\documentclass[aps,prd,preprintnumbers,superscriptaddress,nofootinbib,showpacs]{revtex4}
\usepackage[dvipdfmx]{graphicx}
\usepackage{color} 
\usepackage{amsmath}
\usepackage{amssymb}
\usepackage{bm}
\newcommand{\bea}{\begin{eqnarray}}
\newcommand{\eea}{\end{eqnarray}}
\newcommand{\be}{\begin{equation}}
\newcommand{\ee}{\end{equation}}

 \makeatletter  
\def\alt{\mathrel{\mathpalette\gl@align<}}
\def\agt{\mathrel{\mathpalette\gl@align>}}
\def\gl@align#1#2{ \lower.6ex\vbox{\baselineskip\z@skip\lineskip\z@
\ialign{ $\m@th#1\hfil##\hfil$\crcr#2\crcr\sim\crcr }} } \makeatother

\newcommand{\simgt}{\lower.5ex\hbox{$\; \buildrel > \over \sim \;$}}
\newcommand{\simlt}{\lower.5ex\hbox{$\; \buildrel < \over \sim \;$}}

\begin{document}

\title{
An analytic halo approach to the bispectrum of galaxies in redshift space
}

\author{Kazuhiro Yamamoto}
\affiliation{
Graduate school of Physical Sciences, Hiroshima University, 
Higashi-hiroshima, Kagamiyama 1-3-1, 739-8526, Japan}

\author{Yue Nan}
\affiliation{
Graduate school of Physical Sciences, Hiroshima University, 
Higashi-hiroshima, Kagamiyama 1-3-1, 739-8526, Japan}

\author{Chiaki Hikage}
\affiliation{Kavli Institute for the Physics and Mathematics of the Universe (Kavli IPMU, WPI), \\The University of Tokyo, 5-1-5 Kashiwanoha, Kashiwa, Chiba, 277-8583, Japan}

\begin{abstract}
We present an analytic formula for the galaxy bispectrum in redshift space
on the basis of the halo approach description with the halo occupation 
distribution of central galaxies and satellite galaxies. 
This work is an extension of a previous work on the galaxy power spectrum, 
which illuminated the significant contribution of satellite galaxies to the
higher multipole spectrum through the non-linear redshift space distortions
of their random motions.
Behaviors of the multipoles of the bispectrum are compared with results of 
numerical simulations assuming a halo occupation distribution of 
the LOWZ sample of the SDSS-III BOSS survey. 
Also presented are analytic approximate formulas for the multipoles of the bispectrum, 
which is useful to understanding their characteristic properties.   
We demonstrate that the Fingers of God effect is quite important for the 
higher multipoles of the bispectrum in redshift space, depending on 
the halo occupation distribution parameters. 
\end{abstract}
 
\maketitle

\section{Introduction}
The three-point correlation function and the bispectrum are the simplest 
quantities that characterize the non-Gaussian properties of clustering.
Results from the Planck satellite have shown that the primordial perturbations 
are almost Gaussian \cite{Planckresults}, but non-Gaussian properties
in the density perturbations arise in the course of nonlinear evolution 
of the clustering of matter and galaxies under the effect of gravity \cite{BS1,BS2,BS3,IBispec1,IBispec2}. 
Thus, the galaxy bispectrum is a fundamental tool for characterizing non-Gaussian 
properties of galaxy distributions (for a review, see, e.g., \cite{Scoccimarro2000}). 
A precise galaxy bispectrum was recently measured in the Sloan Digital Sky Survey (SDSS) III baryon oscillation spectroscopic survey (BOSS) galaxies distribution, 
and the usefulness for constraining cosmological parameters was demonstrated 
\cite{HGilMarin1,HGilMarin2}. 
There are many theoretical works on bispectra including redshift space 
distortions (e.g., \cite{scoccimarro1999}).
The bispectrum in modified gravity theories has also been investigated in Refs. 
\cite{Shirata,Koyama,Barreia,Emilio1,Emilio2,Takushima1,Takushima2,IBispec3}.
In general, however, it is difficult to construct a theoretical 
model for galaxy bispectra that fits observational bispectra at small scales, 
even in the framework of Newtonian gravity. 
We challenge this problem to construct an analytic model for the galaxy bispectrum
in redshift space, considering not only monopoles but also higher multipoles of
bispectra, which reflects the redshift space distortions more significantly.

In a previous work \cite{HikageYamamoto}, it was demonstrated that the 
halo approach is quite useful to explain the multipole power spectra in 
redshift space of SDSS luminous red 
galaxies (LRGs), in which the halo occupation distribution (HOD) of 
central galaxies and satellite galaxies plays an important role. 
One halo term in particular makes quite a large contribution to the higher 
multipole spectra at large wavenumbers $k>0.2h^{-1}$Mpc. This discovery 
provides useful applications of the higher multipoles spectrum
in the quasi-linear and nonlinear regimes \cite{Hikage,Kanemaru}.
As an extension of our previous work \cite{HikageYamamoto}, we develop an 
analytic formula for the galaxy bispectrum on the basis of the halo approach 
with the HOD of central and satellite galaxies.
In Ref.~\cite{smith2008}, the authors presented an analytic model of the 
bispectrum of galaxies in redshift space with the halo approach. However, 
our work utilizes a framework with central galaxies and satellite galaxies, 
which plays a crucial role in the theoretical formula. 
In particular, we show that satellite galaxies are essential to accurately 
describe the multipoles of bispectrum in redshift space.

This paper is organized as follows. 
In section 2, we first review the bispectrum in the standard perturbation 
theory as well as the halo approach description of galaxy clustering, 
which are the basis of our theoretical model of the bispectrum. 
Then, in section 3, the multipoles of the bispectrum and the reduced bispectrum 
are introduced. 
The characteristic behaviors of the multipoles of the reduced bispectrum 
are demonstrated by adopting the HOD of the SDSS-II LRG sample and the SDSS-III 
BOSS low redshift (LOWZ) sample. 
The multipole bispectrum is compared with the result of numerical simulations 
by adopting the same HOD of the SDSS-III BOSS LOWZ sample. 
Analytic approximate formulas for the multipoles of the bispectrum are
presented in section 4, which is useful to understand their characteristic properties.   
Section 5 is devoted to summary and conclusions. 
The Appendix lists analytic formulas that are useful for the multipoles 
of bispectrum in redshift space.

\def\NL{}
\section{Derivation of the theoretical formula}

\subsection{Bispectrum in the standard perturbation theory}

We start with reviewing the bispectrum in redshift space in the standard 
cosmological perturbation theory \cite{scoccimarro1999}, which is  
useful as an introduction to the bispectrum. 
The fluid equations in an expanding universe are given by
\begin{eqnarray}
&&\dot \delta(t,{\bm x})+{1\over a}\partial_i[(1+\delta(t,{\bm x}))v^i(t,{\bm x})]=0,
\\
&&\dot v^i(t,{\bm x})+{\dot a\over a}v^i(t,{\bm x})+{1\over a}v^j(t,{\bm x})\partial_jv^i(t,{\bm x})=-{1\over a}\partial_i\psi(t,{\bm x}),
\end{eqnarray}
with the cosmological Poisson equation
\begin{eqnarray}
&&\triangle \psi(t,{\bm x})=4\pi Ga^2\bar \rho_m\delta,
\end{eqnarray}
where $\delta(t,\bm x)$ and $v^i(t,\bm x)$ are the density contrast and the velocity field 
respectively, $\psi$ is the gravitational potential, $a$ is the scale factor,
the dot denotes the differentiation with respect to the cosmic time $t$,
${\bm x}$ is the comoging coordinate, 
and $\bar \rho_m$ is the background matter density. 
The spatially flat Friedmann equation is 
\begin{eqnarray}
&&H^2={\dot a^2\over a^2}={8\pi G\over 3}\bar \rho_m+{\Lambda\over 3},
%
\end{eqnarray}
where $\Lambda$ is the cosmological constant. Using the Hubble parameter $H$, 
the cosmological Poisson equation is written as 
$\triangle \psi(t,{\bm x})={3\over 2}a^2 H^2\Omega_m\delta$, 
where $\Omega_m$ is the density parameter at the present epoch.

As we consider only scalar mode perturbations, by introducing the Fourier expansion 
for $\delta$ and $\theta(=\nabla_i v^i/aH)$, 
\begin{eqnarray}
&&\delta(t,{\bm x})={1\over (2\pi)^3}\int d^3p \delta(t,{\bm p})e^{i{\bm p}\cdot{\bm x}},
\\
&&v^i(t,{\bm x})={1\over (2\pi)^3}\int d^3p {-i p^i\over p^2}aH\theta(t,{\bm p})e^{i{\bm p}\cdot{\bm x}},
\end{eqnarray}
the fluid equations reduce to
\begin{eqnarray}
&&{1\over H}\dot \delta(t,{\bm p})+\theta(t,{\bm p})
=-{1\over (2\pi)^3}\int d^3k_1\int d^3k_2
\delta^{(3)}_D({\bm k}_1+{\bm k}_2-{\bm p})\left(1+{{\bm k}_1\cdot{\bm k}_2\over k_2^2}\right)
\delta(t,{\bm k}_1)\theta(t,{\bm k}_2),
\label{hydroA}
\\
&&{1\over H}\dot \theta(t,{\bm p})+{1\over 2}\theta(t,{\bm p})-{p^2\over a^2H^2}\psi(t,{\bm p})
\nonumber\\
&&\hspace{0.1cm}
=-{1\over (2\pi)^3}\int d^3k_1\int d^3k_2
\delta^{(3)}_D({\bm k}_1+{\bm k}_2-{\bm p})
\left({({\bm k}_1\cdot{\bm k}_2)|{\bm k}_1+{\bm k}_2|^2\over 2k_1^2k_2^2}\right)
\theta(t,{\bm k}_1)\theta(t,{\bm k}_2),
\label{hydroB}
\end{eqnarray}
where $\delta_D^{(3)}({\bm k}_1+{\bm k}_2-{\bm p})$ denotes the Dirac's delta function. 
Following the standard cosmological perturbation theory, we find the solution in the expanded form
(see e.g., Refs.~\cite{Takushima1,Takushima2}), 
\begin{eqnarray}
  &&\delta(t,{\bm p})=D_1(t)\delta_L({\bm p})+D_1^2(t)\left({\cal W}_\alpha({\bm p})-{2\over 7}\lambda(t){\cal W}_\gamma({\bm p})\right)+\cdots,
  \label{deltatpnA}
\\
&&\theta(t,{\bm p})=-f(t)\left[D_1(t)\delta_L({\bm p})+D_1^2(t)\left({\cal W}_\alpha({\bm p})-{4\over 7}\lambda_\theta(t){\cal W}_\gamma({\bm p})\right)+\cdots\right],
\label{deltatvnA}
\end{eqnarray}
where $D_1(t)$ and is the linear growth factor, $f(t)=d\ln D_1(t)/d\ln a(t)$ is the linear growth rate, and $\delta_L({\bm p})$
describes the linear density perturbations, which obeys the Gaussian random distribution with
\begin{eqnarray}
 \langle\delta_L({\bm k}_1) \delta_L({\bm k}_2) 
 \rangle=P_m(k_1)\delta^{(3)}_D({\bm k}_1+{\bm k}_2),
 \label{defofPL}
\end{eqnarray}
where $P_m(k)$ is the matter power spectrum. 
Here ${\cal W}_\alpha({\bm p})$ and ${\cal W}_\gamma({\bm p})$ are defined as
\begin{eqnarray}
  &&  {\cal W}_\alpha({\bm p})={1\over (2\pi)^3}\int d{\bm k}_1\int d{\bm k}_2\delta_D^{(3)}({\bm k}_1+{\bm k}_2-{\bm p})
  \alpha^{(s)}({\bm k}_1,{\bm k}_2)\delta_L({\bm k}_1)\delta_L({\bm k}_2),
  \\
&&  {\cal W}_\alpha({\bm p})={1\over (2\pi)^3}\int d{\bm k}_1\int d{\bm k}_2\delta_D^{(3)}({\bm k}_1+{\bm k}_2-{\bm p})
  \gamma({\bm k}_1,{\bm k}_2)\delta_L({\bm k}_1)\delta_L({\bm k}_2),
  \end{eqnarray}
with
\begin{eqnarray}
&&  \alpha^{(s)}({\bm k}_1,{\bm k}_2)=1+{{\bm k}_1\cdot{\bm k}_2(k_1^2+k_2^2)\over 2k_1^2k_2^2},
  \\
&&   \gamma({\bm k}_1,{\bm k}_2)=1-{({\bm k}_1\cdot{\bm k}_2)^2\over k_1^2k_2^2},
  \end{eqnarray}
$\lambda(t)$ and $\lambda_\theta(t)$ are also defined as
\begin{eqnarray}
  &&\lambda(t)={7\over 2D_1^2(t)}\int_0^tdt'{D_2(t)D_1(t')-D_1(t)D_2(t')\over D_1(t')\dot D_2(t')-\dot D_1(t')D_2(t')}
  D_1^2(t')f^2(t')H^2(t'),
  \\
&&\lambda_\theta(t)=\lambda(t)+{\dot \lambda(t)\over 2f(t)H(t)},
\end{eqnarray}
where $D_2(t)$ is the decaying mode solution. Note that $\lambda(t)$ and $\lambda_\theta(t)$ reduce
to one in the limit of the Einstein de Sitter universe (see also \cite{SS,JB}).
In the present paper, we adopt an approximation $\lambda(t)=\lambda_\theta(t)=1$,
whose validity is shown in Refs.\cite{Goroff,Takushima1,Takushima2}, then, we write the
solution up to the second order as,
\begin{eqnarray}
  &&\delta(t,{\bm p})=D_1(t)\delta_L({\bm p})+{D_1^2(t)\over (2\pi)^3}
\int d{\bm k}_1\int d{\bm k}_2\delta_D^{(3)}({\bm k}_1+{\bm k}_2-{\bm p})
F_{2}({\bm k}_1,{\bm k}_2)\delta_L({\bm k}_1)\delta_L({\bm k}_2)
  +\cdots,
  \label{deltatpn}
\\
&&\theta(t,{\bm p})=-f(t)\left[D_1(t)\delta_L({\bm p})+
  {D_1^2(t)\over (2\pi)^3}
\int d{\bm k}_1\int d{\bm k}_2\delta_D^{(3)}({\bm k}_1+{\bm k}_2-{\bm p})
G_{2}({\bm k}_1,{\bm k}_2)\delta_L({\bm k}_1)\delta_L({\bm k}_2)
  +\cdots\right],
\label{deltatvn}
\end{eqnarray}
where we defined
\begin{eqnarray}
&&F_2({\bm k}_1,{\bm k}_2)=\alpha^{(s)}({\bm k}_1,{\bm k}_2)
-{2\over 7}\gamma({\bm k}_1,{\bm k}_2),
\label{F2kk}
\\
&&G_2({\bm k}_1,{\bm k}_2)=\alpha^{(s)}({\bm k}_1,{\bm k}_2)
-{4\over 7}\gamma({\bm k}_1,{\bm k}_2).
\label{G2kk}
\end{eqnarray}

Now we consider the density contrast in the redshift space, and define
\begin{eqnarray}
&&u^i=-{v^i\over aHf},
\\
&&{\bm s}={\bm x}-{\bm \gamma}f(\bm \gamma\cdot \bm u),
\end{eqnarray}
where ${\bm s}$ denotes the coordinates in the redshift space,
${\bm \gamma}$ is the line of sight direction.
Hereafter, we write the linear growth rate as $f=d\ln D_1(a)/d\ln a $. 
The Fourier coefficient of the density contrast in redshift space $\delta^s(t,{\bm s})$ 
is defined by 
\begin{eqnarray}
\delta^s(t,{\bm p})&=&\int{d^3s\over (2\pi)^3}\delta^s(t,{\bm s})e^{-i{\bm p}\cdot{\bm s}}
\nonumber\\
&=&\int{d^3x\over (2\pi)^3}e^{-i{\bm p}\cdot{\bm x}}\left(
\delta(t,{\bm x})+f\bm\gamma\cdot\bm\nabla(\bm \gamma\cdot \bm u)\right)
e^{if(\bm \gamma\cdot \bm u)(\bm \gamma\cdot \bm p)}.
\end{eqnarray}
By using the expansion, 
\begin{eqnarray}
e^{if(\bm \gamma\cdot \bm u)(\bm \gamma\cdot \bm p)}
  &=&
\sum_{n=0}{\bigl(if(\bm \gamma\cdot \bm u) (\bm \gamma\cdot \bm p)\bigr)^n\over n!},
\end{eqnarray}
up to the second order of perturbations, we have
\begin{eqnarray}
&&\delta^s(t,{\bm p})=\delta(t,{\bm p})-f\mu^2\theta(t,{\bm p})
-\int d^3k_1\int d^3k_2\delta^{(3)}_D({\bm k}_1+{\bm k}_2-{\bm p})(\delta(t,{\bm k}_1)+f\mu_1^2\theta(t,{\bm k}_1))
f\mu p{\mu_2\over k_2}\theta(t,{\bm k}_2),
\end{eqnarray}
where we define
\begin{eqnarray}
&&\mu={{\bm \gamma}\cdot {\bm p}\over p},
\\
&&\mu_i={{\bm \gamma}\cdot {\bm k}_i\over k_i}.
\end{eqnarray}

By assuming that the galaxy density contrast $\delta_g(t,\bm x)$ is related to the matter 
density contrast  $\delta(t,\bm x)$ as
\begin{eqnarray}
\delta_g(t,{\bm x})&=&b\delta(t,{\bm x})+{b_2\over 2}\delta^2(t,{\bm x}),
\end{eqnarray}
where $b$ and $b_2$ are the constants, 
the Fourier coefficient of the galaxy density contrast in redshift space is written as
\begin{eqnarray}
\delta_g^s(t,{\bm p})&=&b\delta(t,{\bm p})-f\mu^2\theta(t,{\bm p})
-{1\over (2\pi)^3}
\int d^3k_1\int d^3k_2\delta^{(3)}_D({\bm k}_1+{\bm k}_2-{\bm p})(b\delta(t,{\bm k}_1)+f\mu_1^2\theta(t,{\bm k}_1))
f\mu p{\mu_2\over k_2}\theta(t,{\bm k}_2)
\nonumber
\\
&&+{1\over (2\pi)^3}
\int d^3k_1\int d^3k_2\delta^{(3)}_D({\bm k}_1+{\bm k}_2-{\bm p}){b_2\over 2}
\delta(t,{\bm k}_1)\delta(t,{\bm k}_2).
\end{eqnarray}
Using the expressions up to the second order of perturbations,
\begin{eqnarray}
&&\delta(t,{\bm p})=D_1(t)\delta_L({\bm p})+{{D_1^2(t)}\over (2\pi)^3}
\int d^3k_1\int d^3k_2\delta^{(3)}_D({\bm k}_1+{\bm k}_2-{\bm p})
F_2({\bm k}_1,{\bm k}_2)\delta_L({\bm k}_1)\delta_L({\bm k}_2),
\\
&&\theta(t,{\bm p})=-D_1(t)\delta_L({\bm p})-{{D_1^2(t)}\over (2\pi)^3}
\int d^3k_1\int d^3k_2\delta^{(3)}_D({\bm k}_1+{\bm k}_2-{\bm p})
G_2({\bm k}_1,{\bm k}_2)\delta_L({\bm k}_1)\delta_L({\bm k}_2),
\end{eqnarray}
the galaxy density contrast in redshift space is
\begin{eqnarray}
&&\delta_g^s(t,{\bm p})=b\biggl(D_1(t)\delta_L({\bm p})+{{D_1^2(t)}\over (2\pi)^3}
\int d^3k_1\int d^3k_2\delta^{(3)}_D({\bm k}_1+{\bm k}_2-{\bm p})
F_2({\bm k}_1,{\bm k}_2)\delta_L({\bm k}_1)\delta_L({\bm k}_2)\biggr)
\nonumber
\\
&&~~~~~~+f\mu^2
\biggl(D_1(t)\delta_L({\bm p})+{{D_1^2(t)}\over (2\pi)^3}
\int d^3k_1\int d^3k_2\delta^{(3)}_D({\bm k}_1+{\bm k}_2-{\bm p})
G_2({\bm k}_1,{\bm k}_2)\delta_L({\bm k}_1)\delta_L({\bm k}_2)\biggr)
\nonumber
\\
&&~~~~~~
+{{D_1^2(t)}\over (2\pi)^3}\int d^3k_1\int d^3k_2\delta^{(3)}_D({\bm k}_1+{\bm k}_2-{\bm p})(b\delta_L({\bm k}_1)+f\mu_1^2\delta_L({\bm k}_1))
f\mu p{\mu_2\over k_2}\delta_L({\bm k}_2)
\nonumber
\\
&&~~~~~~
+{{D_1^2(t)}\over (2\pi)^3}\int d^3k_1\int d^3k_2\delta^{(3)}_D({\bm k}_1+{\bm k}_2-{\bm p})
{b_2\over 2}\delta_L({\bm k}_1)\delta_L({\bm k}_2).
\end{eqnarray}

Next we compute the three-point clustering statistics. 
With the use of the relation
\begin{eqnarray}
&&\langle \delta_L({\bm p}_1) \delta_L({\bm p}_2) \delta_L({\bm p}_3) \delta_L({\bm p}_4) 
\rangle=\langle \delta_L({\bm p}_1) \delta_L({\bm p}_2)\rangle\langle \delta_L({\bm p}_3) \delta_L({\bm p}_4) \rangle
+\langle \delta_L({\bm p}_1) \delta_L({\bm p}_3)\rangle\langle \delta_L({\bm p}_2) 
\delta_L({\bm p}_4)\rangle
\nonumber
\\
&&
~~~~~~~~~~~~~~~~~~~~~~~~~~~~~~~~~~~~~~
+\langle \delta_L({\bm p}_1) \delta_L({\bm p}_4)\rangle\langle \delta_L({\bm p}_2) \delta_L({\bm p}_3) 
\rangle
\end{eqnarray}
and Eq. (\ref{defofPL}), we have
\begin{eqnarray}
\langle \delta_g^s(t,{\bm k}_1) \delta_g^s(t,{\bm k}_2) \delta_g^s(t,{\bm k}_3) \rangle=(2\pi)^3\delta^{(3)}_D
({\bm k}_1+{\bm k}_2+{\bm k}_3)D_1^4(t)B({\bm k}_1,{\bm k}_2,{\bm k}_3)
\end{eqnarray}
where we defined
\begin{eqnarray}
B({\bm k}_1,{\bm k}_2,{\bm k}_3)=2Z_1({\bm k}_1)Z_1({\bm k}_2)Z_2({\bm k}_1,{\bm k}_2)P_m(k_1)P_m(k_2)
+2~{\rm cyclic~terms},
\end{eqnarray}
and
\begin{eqnarray}
&&Z_1({\bm k})=b+f\mu^2,
\\
&&Z_2({\bm k}_1,{\bm k}_2)=bF_2({\bm k}_1,{\bm k}_2)+f\mu_{12}^2G_2({\bm k}_1,{\bm k}_2)+{b_2\over 2}+{f\mu_{12} k_{12}\over 2}
\left({\mu_2\over k_2}(b+f\mu_1^2)+{\mu_1\over k_1}(b+f\mu_2^2)\right),
\end{eqnarray}
and $\mu={\bm k}\cdot \bm{\gamma}/k$, $\mu_i={\bm k}_i\cdot \bm{\gamma}/k_i$, $k_i=|{\bm k}_i|$, 
$\mu_{12}=({\bm k}_1+{\bm k}_2)\cdot \bm{\gamma}/k_{12}$, and $k_{12}=|{\bm k}_1+{\bm k}_2|$.
We may also rewrite (\ref{F2kk}) and (\ref{G2kk}) as
\begin{eqnarray}
&&F_2({\bm k}_i,{\bm k}_j)={5\over 7}+{x_{ij}\over 2}\left({k_i\over k_j}+{k_j\over k_i}\right)+{2x_{ij}^2\over 7},
\\
&&G_2({\bm k}_i,{\bm k}_j)={3\over 7}+{x_{ij}\over 2}\left({k_i\over k_j}+{k_j\over k_i}\right)+{4x_{ij}^2\over 7},
\end{eqnarray}
where we use the notation $x_{ij}=\cos\theta_{ij}={\bm k}_i\cdot{\bm k}_j/k_ik_j$.

\begin{figure}[t]
\begin{center}
\vspace{0cm}
\includegraphics[width=120mm]{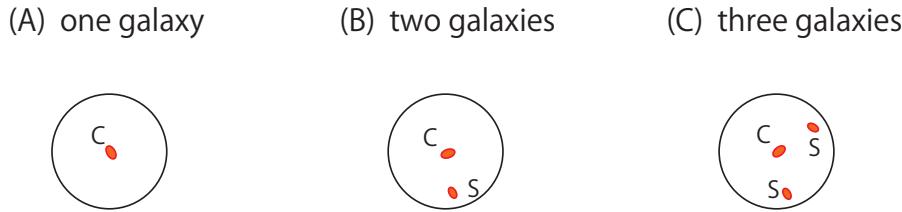}
\vspace{0cm}
\end{center}
\caption{Case analysis of galaxy distribution in halos. 
(A) Case where one galaxy in a halo is a central galaxy in one halo. 
(B) Case where two galaxies in a halo are a pair of a central galaxy and a satellite galaxy 
in one halo. 
(C) Case where three galaxies in a halo are a combination of  
a central galaxy and two satellite galaxies in one halo.
}
\label{fig:galaxiesOH}
\end{figure}

\subsection{Halo approach} 
The halo approach is useful to describe distributions of dark matter 
as well as distributions of galaxies from large to small scales 
\cite{White,Seljak,Scoccimarro2001,CooraySheth2002}. 
In this approach, all dark matter and galaxies are associated with 
virialized dark matter halos. The basic quantities of this approach 
are the dark matter density profile of a halo $\rho(r)$ and the halo mass 
function $dn/dM$. 

In the present paper, we assume the Navarro--Frenk--White (NFW) density profile \cite{NFWprofile},
\begin{eqnarray}
\rho(r)={\rho_s\over r/r_s(1+r/r_s)^2},
\end{eqnarray}
where the characteristic density $\rho_s$ 
and characteristic scale $r_s$ are fitting parameters.
The virial mass of a halo within the virial radius $r_{\rm vir}$, 
which is related to the concentration parameter $c$ by $c=r_{\rm vir}/r_s$, 
is defined such that the averaged density within the radius is 
$\Delta_{\rm vir}$ times of the mean matter density $\bar \rho_m(z)$. 
Then, the virial mass $M_{\rm vir}$ is written as 
\begin{eqnarray}
M_{\rm vir}=4\pi \int_0^{r_{\rm vir}}dr r^2\rho(r)=4\pi \rho_sr_s^3\left(
\ln(1+r/r_s)-{r/r_s\over 1+r/r_s}\right)={4\pi\over 3}r_{\rm vir}^3\Delta_{\rm vir}(z)
\bar \rho_m(z),
\end{eqnarray}
where we adopt $\Delta_{\rm vir}=265$ at $z=0.3$. 
We use $M_{\rm vir}$ as the mass of halos. 
Then, we introduce the
Fourier transform of the truncated NFW profile (see \cite{Scoccimarro2001,CooraySheth2002}),
\begin{eqnarray}
\tilde u_{\rm NFW}(k;M)&=&{\int_{r\leq r_{\rm vir}}d^3x \rho(r) e^{-i{\bm k}\cdot{\bm x}}\over
\int_{r\leq r_{\rm vir}}d^3x \rho(r)}
\nonumber
\\
&=&{4\pi \rho_s r_s^3\over M}
\biggl\{ \sin (kr_s) \left[ Si([1+c]kr_s)-Si(kr_s)\right] \
-{\sin ckr_s\over(1+c)kr_s}
+\cos (kr_s) \left[ Ci([1+c]kr_s)-Ci(kr_s)\right]\biggr\}, 
\nonumber\\
\label{tuNFW}
\end{eqnarray}
where $C_i(x)$ and $S_i(x)$ are defined by
\begin{eqnarray}
C_i(x)=-\int_x^\infty {\cos t\over t} dt,~~~~~~
S_i(x)=\int_0^x {\sin t\over t} dt.~~~~~~
\end{eqnarray}

Because we are interested in the distribution of galaxies, we introduce the 
halo occupation distribution $N_{\rm  HOD}(M)$, which describes the average 
number of galaxies inside a halo with mass $M$. 
In our approach, we introduce a description with central and 
satellite galaxies. Central galaxies reside at the centers of halos, 
while satellite galaxies reside in off-center regions of halos with large random 
velocities. We use the following form of the HOD with central galaxies and 
satellite galaxies \cite{Zheng2005},
\begin{eqnarray}
&&N_{\rm HOD}(M)=\langle N_{\rm cen}\rangle(1+\langle N_{\rm sat}\rangle), \\
&&\langle N_{\rm cen}\rangle =\frac{1}{2}\left[1+{\rm erf}\left(\frac{\log_{10}(M)-\log_{10}
(M_{\rm min})}{\sigma_{\log M}}\right)\right], \\
&&\langle N_{\rm sat}\rangle =
\left(\frac{M-M_{\rm cut}}{M_1}\right)^{\alpha},
\label{eq:HOD}
\end{eqnarray}
where ${\rm erf}(x)$ is the error function.
We adopt the HOD parameters listed in Table I
for the SDSS LRG catalog \cite{ReidSpergel} and for the 
SDSS-III BOSS LOWZ catalog \cite{Parejko}.
Assuming that the number of groups with $N_{\rm sat}$ satellites follows the Poisson
distribution \cite{Kravtsov2004}, the averaged satellite--satellite pair
number $\langle N_{\rm sat}(N_{\rm sat}-1)\rangle$ per halo goes to
$\langle N_{\rm cen}\rangle\langle N_{\rm sat}\rangle^2$.  
A deviation from the Poisson distribution for smaller halos could be influential 
for estimating the second and third moments of the galaxy distribution within a 
halo. We have checked the effect by introducing the formulas
$\langle N_{\rm sat}(N_{\rm sat}-1) \rangle=\alpha^2(M)\langle N_{\rm sat}\rangle^2$
and $\langle N_{\rm sat}(N_{\rm sat}-1)(N_{\rm sat}-2) \rangle=\alpha^3(M)\langle N_{\rm sat}\rangle^3$
with $\alpha(M)$ defined in Refs. \cite{Scranton2002,TakadaJain,Scoccimarro2001a}
\begin{eqnarray}
\alpha(M)=\left\{
\begin{array}{ll}
1 & ~{\rm for}~M\geq 10^{13} h^{-1} M_\odot,\\
\ln \sqrt{M/10^{11}h^{-1}M_\odot} & ~{\rm for}~M< 10^{13} h^{-1} M_\odot.\\
\end{array}
\right.
\end{eqnarray}
However, this effect does not alter our results because the halo mass of 
the galaxy samples adopted in the present work is large. 

\begin{table}[t]
\begin{center}
\begin{tabular}{cccc}
\hline
\hline
~ & ~~~~~~~~~~~~~LRG~~~~~~~~~~~~~& ~~~~~~~~~~~~~LOWZ ~~~~~~~~~~~~~ \\
\hline
$M_{\rm min}$ & $5.7\times 10^{13}h^{-1}M_\odot$ & $1.5\times 10^{13}h^{-1}M_\odot$ \\
$\sigma_{\log M}$ & 0.7 & 0.45  \\
$M_{\rm  cut}$ & $3.5\times 10^{13}h^{-1}M_\odot$  &  $1.4\times 10^{13}h^{-1}M_\odot$ \\
$M_1$ & $3.5\times10^{14}h^{-1}M_\odot$ & $1.3\times10^{14}h^{-1}M_\odot$  \\
$\alpha$ & $1$ & $1.38$ \\
\hline\hline
\end{tabular}
\caption{HOD parameters for the LRG samples~\cite{ReidSpergel}  and the LOWZ sample \cite{Parejko}.}
\end{center}
\label{tab:lrg_halo}
\end{table}

We assume that the distribution of satellite galaxies follows 
the NFW profile, and that the Fourier transform of the truncated NFW 
profile (\ref{tuNFW}) represents the power spectrum of the one-halo term. 
These assumptions do not alter our results. 
We assume that the satellite galaxies have internal random velocities following 
a Gaussian distribution specified by the one-dimensional velocity dispersion 
\cite{HikageYamamoto,Kanemaru,ReidSpergel,LokasMamon},
\begin{eqnarray}
\sigma_{v,{\rm off}}(M)=\left({GM\over2r_{\rm vir}}\right)^{1/2}.
\label{sigmavoff}
\end{eqnarray}
These random motions cause the Fingers of God (FoG) effect,
which changes the distribution of satellite galaxies in redshift space.
Assuming that satellite motions in a halo are uncorrelated with each other, 
then the Fourier transform of the distribution of the satellite 
galaxies in redshift space is obtained (e.g., \cite{HikageYamamoto}) as
\begin{eqnarray}
\label{eq:psoff}
\widetilde u(\bm k,M)&=&\tilde{u}_{\rm NFW}(k;M)\exp\left[-\frac{\sigma_{v,{\rm off}}^2(M)k^2\mu^2}{2a^2H^2(z)}\right].
\label{eq:psoff2}
\end{eqnarray}

The other key quantity of the halo-approach description is the halo mass function $dn/dM$, 
which is the number density of halos with mass $M$ per unit volume and per unit mass. 
For the halo mass function, we adopt the fitting formula in Refs.~\cite{ShethTormen1999,ShethTormen2002,Parkinson2007}
\begin{eqnarray}
 M{dn\over dM}={\bar \rho_m \over M}{d\ln \sigma_R^{-1}\over d\ln M} f(\sigma_R)
\end{eqnarray}
with
\begin{eqnarray}
 f(\sigma_R)=0.322\sqrt{2\times0.707\over \pi}\left[
1+\left({1\over 0.707\nu^2}\right)^{0.3}\right]{\nu}\exp
\left(-{0.707\nu^2\over 2}\right),
\label{massfunctionmodel}
\end{eqnarray}
and  $\nu = \delta_c/\sigma(R)$,
where $\sigma_R$ is the root mean square fluctuation in spheres containing
 mass $M$ at the initial time, which is extrapolated to the redshift $z$ using linear theory, 
$\delta_c$ is the critical value of the initial overdensity 
which is required for collapse, and $\delta_c=1.686$ is adopted.

For the linear bias $b(M)$, we adopt the halo bias of the fitting function,
\begin{eqnarray}
b(M)=1-{\nu^a\over \nu^a+\delta_c^a}+0.183\nu^b+0.265 \nu^c
\label{biasmodel}
\end{eqnarray}
with $a = 0.132$, $b = 1.5$ and $c = 2.4$, which was calibrated using N-body simulations 
\cite{Tinker}.

The power spectrum in the halo approach is given by 
a combination of the one-halo term and the two-halo term (see \cite{HikageYamamoto}),
\begin{eqnarray}
P_g(t,\bm k)=P_{g,1h}(t,{\bm k})+P_{g,2h}(t,{\bm k}).
\label{Pgstk}
\end{eqnarray}
The one-halo term is given by 
\begin{eqnarray}
&&P_{g,1h}(t,{\bm k})={1\over {\bar n}^2}\int dM {dn\over dM}
\left[2\left<N_c\right>\left<N_s\right>\widetilde u({\bm k},M)
+\left<N_s(N_s-1)\right>\widetilde u^2({\bm k},M)\right],
\label{onehaloterm}
\end{eqnarray}
where $\bar{n}$ is the mean number density of galaxies given by 
\begin{eqnarray}
\bar{n}=\int dM {dn\over dM} N_{\rm HOD}(M), 
\end{eqnarray}
while the two-halo term is given by
\begin{eqnarray}
&&P_{g,2h}(t,{\bm k})={1\over {\bar n}^2}\prod_{i=1}^2\left[\int dM_i {dn\over dM_i}
\left<N_c\right>\left\{1+\left<N_s\right>\widetilde u({\bm k},M)\right\}(b(M_i)+f\mu^2)
\right]P_m^{\NL}(t,k),
\label{twohaloterm}
\end{eqnarray}
where $P_m^{\NL}(t,k)$ is the matter power spectrum at the time $t$, 
for which we use the
nonlinear fitting formula for the matter power spectrum \cite{NFWprofile}. 
We also use the fitting formula of the linear growth rate
$f={d\log D_1(a)/d\log a}=[\Omega_m(a)]^\gamma$,
where $\Omega_m(a)$ is the matter density parameter at the scale
factor $a=a(t)$ and $\gamma=0.55$.

The one-halo term of the power spectrum (\ref{onehaloterm}) represents 
the contribution from a pair of galaxies in one halo, as is shown in
panel (B) of Fig.~\ref{fig:galaxiesOH}. 
The two-halo term (\ref{twohaloterm}) 
represents the contribution from a pair of galaxies in two different 
halos, as are obtained by combinations of the galaxies in
each panel of Fig.~\ref{fig:galaxiesOH}. 

In Ref.~\cite{HikageYamamoto}, using the SDSS LRG sample, 
the authors demonstrated that satellite galaxies make a significant contribution 
to the multipole power spectrum even when their fraction is small, 
where the multipole galaxy power spectrum of $P_g(t,\bm k)$ is defined by 
\begin{eqnarray}
P^\ell(t,k)=\int_{-1}^{+1}d\mu P_{g}(t,{\bm k}){\cal L}_\ell(\mu),
\end{eqnarray}
using the Legendre polynomial ${\cal L}_\ell(\mu)$ with $\mu=\bm \gamma\cdot \bm k/k$.
The one-halo term, describing the FoG effect of satellite galaxies, 
makes the dominant contribution to the higher multipole spectra with $\ell\geq2$. 
It is also demonstrated that small-scale information of the higher multipole spectrum is 
useful for calibrating the satellite FoG effect and testing gravity theory on halo scales 
\cite{Kanemaru} and
dramatically improves the measurement of the cosmic 
growth rate \cite{Hikage}.

\begin{figure}[t]
\begin{center}
\includegraphics[width=140mm]{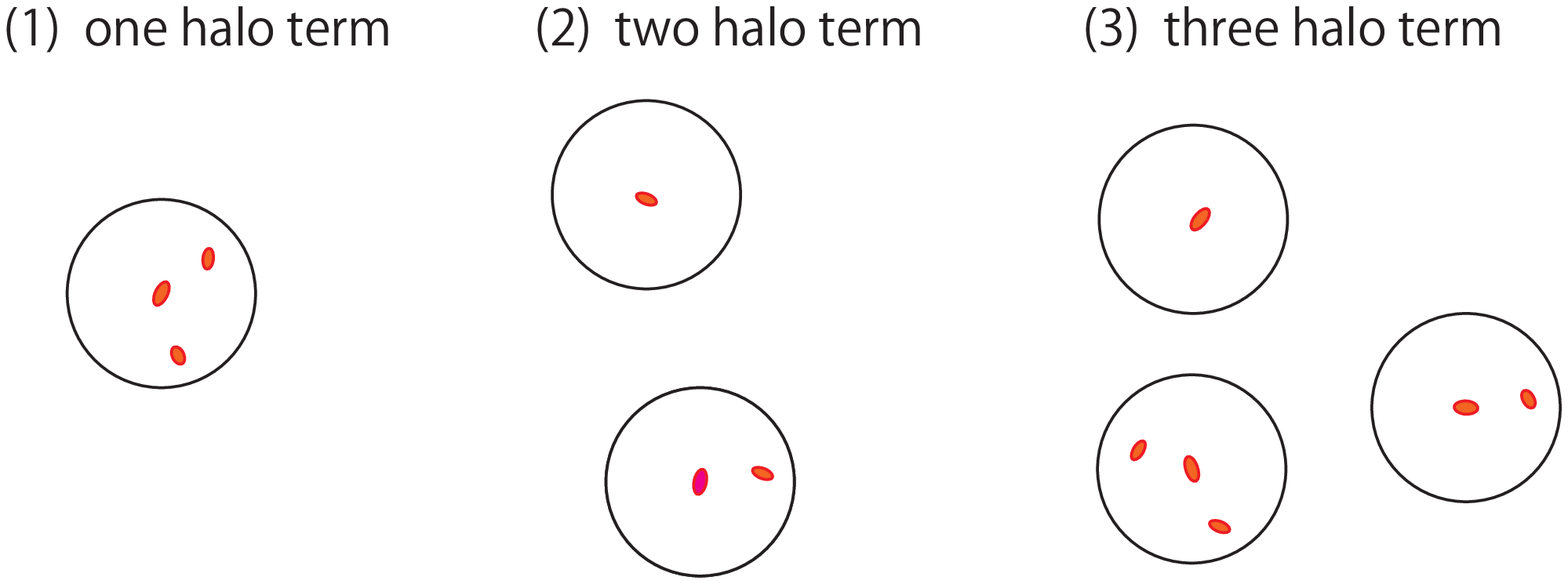}
\caption{Contributions of a (1)  one-halo term, (2) two-halo term, and (3) three-halo 
term to the bispectrum. 
(1) A one-halo term represents the contribution from three galaxies in one halo.
(2) A two-halo term represents the contribution from the combination of two halos with one galaxy and 
two galaxies in each halo. 
(3) A three-halo term represents the contribution from three halos with each galaxy. 
}
\label{fig:galaxiesH}
\end{center}
\end{figure}

\section{Bispectrum in the halo approach}
\begin{figure}[t]
\begin{center}
\includegraphics[width=80mm]{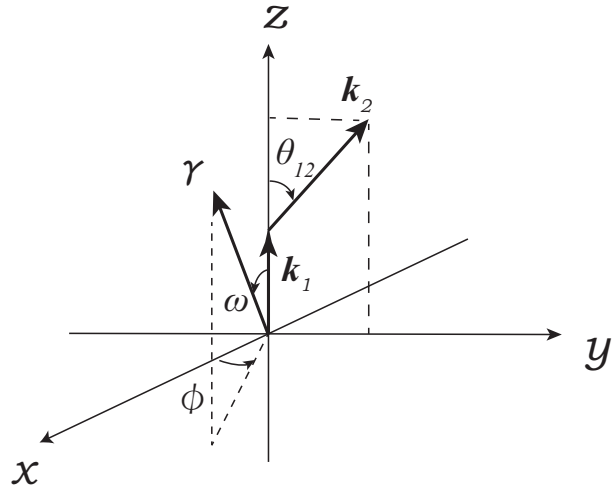}
\caption{Definition of variables for the bispectrum (see also Table II).} 
\label{fig:configuration}
\end{center}
\end{figure}
\begin{table}[h]
\begin{center}
\begin{tabular}{cl}
\hline
\hline
~~~~{\rm Variables}~~~~~~~ & ~~~~~~~~~~~~~~~~~~~~~~{\rm Meaning} \\
\hline
$k_i$ & Magnitude of the wavenumber vector $\bm k_i$\\
$\hat {\bm k}_i$ & Unit vector of the wavenumber vector $\bm k_i$\\
$\mu_i(=\hat {\bm k}_i\cdot\bm \gamma)$ & Cosine of the angle between 
$\bm k_i$ and the line of sight direction $\bm \gamma$ \\
$\mu(=\mu_1)$ & Cosine of the angle ($\omega$) between 
$\bm k_1$ and the line of sight direction $\bm \gamma$ \\
$\theta_{12}(=\theta)$ & Angle between $\bm k_1$ and $\bm k_2$\\
$\omega$ & Angle between $\bm k_1$ and the line of sight direction $\bm \gamma$\\
$\phi$ & Azimuthal angle of the line of sight direction $\bm \gamma$ around $\bm k_1$\\
\hline\hline
\end{tabular}
\caption{
Definition of the variables for the bispectrum (see also Figure 3).
The bispectrum is the function of the five variables, $k_1$, $k_2$, $\theta$, 
$\omega$, and $\phi$.
}
\end{center}
\label{DefVariables}
\end{table}

\subsection{Variables for the bispectrum in redshift space}

The bispectrum with the halo approach is investigated in Ref.~\cite{smith2008}. 
In the present paper, as a generalization of a previous work \cite{smith2008}, 
we present an analytic expression for the bispectrum applying the halo approach 
with the HOD description with central galaxies and satellite galaxies. 
We focus on the bispectrum $B_g(t,\bm k_1,\bm k_2, \bm k_3)$, which is defined by 
\begin{eqnarray}
\left\langle\delta(t,\bm k_1)\delta(t,\bm k_3)\delta(t,\bm k_3)\right\rangle
=(2\pi)^3\delta_D^{(3)}(\bm k_1+\bm k_2+\bm k_3)B_g(t,\bm k_1,\bm k_2, \bm k_3). 
\end{eqnarray}
Thus the bispectrum $B_g(t,\bm k_1,\bm k_2, \bm k_3)$ 
implicitly assumes ${\bm k}_1+{\bm k}_2+{\bm k}_3=0$. 
Since we have the constraint ${\bm k}_1+{\bm k}_2+{\bm k}_3=0$, 
the bispectrum is described by the five parameters, $k_1$, $k_2$, $\cos\theta_{12}(={{\bm k}}_1\cdot{\bm k}_2/k_1k_2)$, 
$\mu(=\cos\omega)$, and $\phi$, as variables of the bispectrum,
with which we write the vectors
\begin{eqnarray}
&&{\bm k}_1=(0,0,k_1),
\\
&&{\bm k}_2=(0,k_2\sin\theta_{12},k_2\cos\theta_{12}),
\\
&&{\bm k}_3=(0,-k_2\sin\theta_{12},-k_1-k_2\cos\theta_{12}),
\\
&&{\bm \gamma}=(\sin\omega\cos\phi,\sin\omega\sin\phi,\cos\omega). 
\end{eqnarray}
For the configuration of the variables, see Fig.~\ref{fig:configuration} and Table II. 
Then, we can write $\mu_i$ as
\begin{eqnarray}
&&\mu_1={\hat{\bm k}}_1\cdot {\bm \gamma}=\cos\omega=\mu,
\\
&&\mu_2={\hat{\bm k}}_2\cdot {\bm \gamma}=
\sin\theta_{12}\sin\omega\sin\phi
+\cos\theta_{12}\cos\omega
=\sin\theta_{12}\sqrt{1-\mu^2}\sin\phi
+\cos\theta_{12}\mu,
\\
&&\mu_3=-{k_1\over k_3}\mu_1-{k_2\over k_3}\mu_2,
\end{eqnarray}
with ${k}_3^2=(k_2\sin\theta_{12})^2+(k_1+k_2\cos\theta_{12})^2$. 
Hereafter, we use the notation $\theta=\theta_{12}$. 
\def\thetaot{{\theta}}

\subsection{Expression for the bispectrum in redshift space}

The bispectrum in the halo approach consists of the one-halo term $B_{g,1h}$, the two-halo term $B_{g,2h}$, 
and the three-halo term $B_{g,3h}$ as
\begin{eqnarray}
B_g(t,\bm k_1,\bm k_2, \bm k_3)=B_{g,1h}(t,{\bm k}_1,{\bm k}_2,{\bm k}_3)
+B_{g,2h}(t,{\bm k}_1,{\bm k}_2,{\bm k}_3)
+B_{g,3h}(t,{\bm k}_1,{\bm k}_2,{\bm k}_3),
\end{eqnarray}
which are written as
\begin{eqnarray}
B_{g,1h}(t,{\bm k}_1,{\bm k}_2,{\bm k}_3)&=&{1\over \bar n^3}\int dM {dn(M)\over dM} \biggl[
\left<N_c\right>\left<N_s(N_s-1)\right>\left(\widetilde u({\bm k}_1,M)\widetilde u({\bm k}_2,M)
+2~{\rm cyclic~terms}\right)
\nonumber
\\
&&
+\left<N_s(N_s-1)(N_s-2)\right>\widetilde u({\bm k}_1,M)\widetilde u({\bm k}_2,M)\widetilde u({\bm k}_3,M)
\biggr],
\\
B_{g,2h}(t,{\bm k}_1,{\bm k}_2,{\bm k}_3)&=&{1\over \bar n^3}\int dM_1 {dn(M_1)\over dM_1} \biggl[\left<N_c\right>\left<N_s\right>\left(\widetilde u({\bm k}_1,M_1)+\widetilde u({\bm k}_2,M_1)\right)
+\left<N_s(N_s-1)\right>\widetilde u({\bm k}_1,M_1)\widetilde u({\bm k}_2,M_1)\biggr]
\nonumber
\\
&&
\times\int dM_2 {dn(M_2)\over dM_2} \left(
\left<N_c\right>+\left<N_c\right>\left<N_s\right>\widetilde u({\bm k}_3,M_2)\right)P_{2h}(t,{\bm k}_3,M_1,M_2)
+2~{\rm cyclic~terms},
\\
B_{g,3h}(t,{\bm k}_1,{\bm k}_2,{\bm k}_3)
&=&{1\over \bar n^3}\int \prod_{i=1}^3\left[dM_i {dn(M_i)\over dM_i} 
\left<N_c\right>\left(1+\left<N_s\right>\widetilde u({\bm k}_i,M_i)\right)\right]
P_{3h}(t,{\bm k}_1,{\bm k}_2,{\bm k}_3,M_1,M_2,M_3),
\end{eqnarray}
where we define
\begin{eqnarray}
&&\hspace{-1cm}P_{2h}(t,{\bm k}_3,M_1,M_2)=(b(M_1)+\mu_3^2 f)(b(M_2)+\mu_3^2 f)P_{m}^{\rm NL}(t,k_3),
\label{twohalom}
\\
&&\hspace{-1cm}P_{3h}(t,{\bm k}_1,{\bm k}_2,{\bm k}_3,M_1,M_2,M_3)
=2P_{m}^{\NL}(t,k_1)P_{m}^{\NL}(t,k_2)Z_1^{}({\bm k}_1,M_1)Z_1^{}({\bm k}_2,M_2)
Z_2^{}({\bm k}_1,{\bm k}_2,M_3)
+2~{\rm cyclic~terms}
\label{threehalom}
\end{eqnarray}
with
\begin{eqnarray}
&&\hspace{-1cm}Z_1^{}({\bm k}_1,M_1)=b(M_1)+f\mu_1^2,
\\
&&\hspace{-1cm}Z_1^{}({\bm k}_2,M_2)=b(M_2)+f\mu_2^2, 
\\
&&\hspace{-1cm}Z_2^{}({\bm k}_1,{\bm k}_2,M_3)=b(M_3)F_2({\bm k}_1,{\bm k}_2)+{b_2(M_3)\over 2}
+f\mu_{12}^2G_2({\bm k}_1,{\bm k}_2)
\nonumber\\
&&~~~~~~~~~~~~~
+{1\over2}f\mu_{12}k_{12}\left\{{\mu_1\over k_1}\left(b(M_3)+f\mu_2^2\right)
+{\mu_2\over k_2}\left(b(M_3)+f\mu_1^2\right)\right\},
\end{eqnarray}
$\mu_{12}=({\bm k}_1+{\bm k}_2)\cdot {\bm \gamma}/k_{12}$, and $k_{12}=|{\bm k}_1+{\bm k}_2|$. 
The directional cosine between the vector ${\bm k}_j$
and the line-of-sight direction is $\mu_j={\bm k}_j\cdot{{\bm \gamma}}/k_j$. 
Hereafter, we set $b_2=0$ unless otherwise stated explicitly.


\begin{figure}[t]
\begin{center}
\includegraphics[width=79mm]{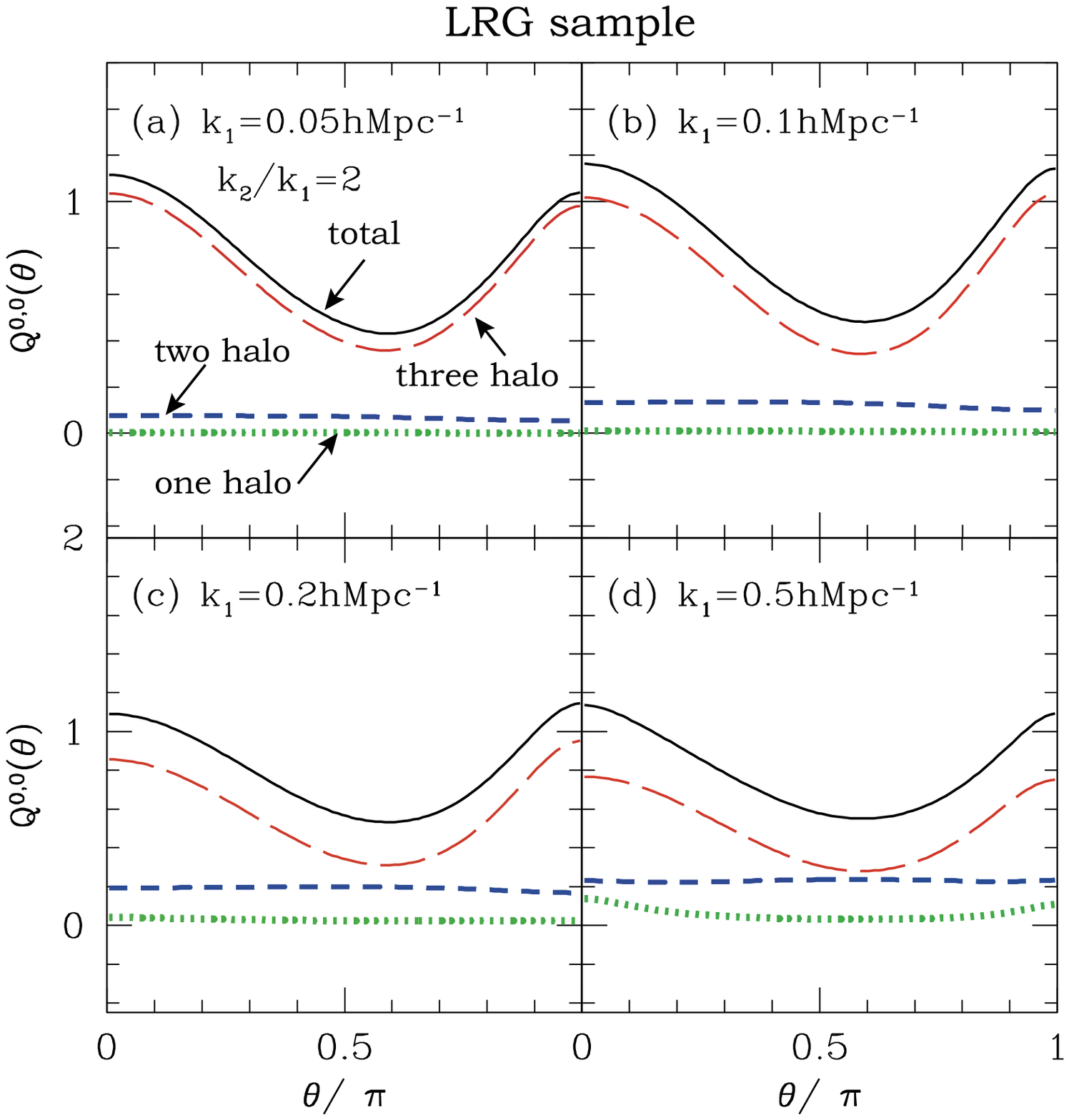}
\hspace{0.9cm}
\includegraphics[width=79mm]{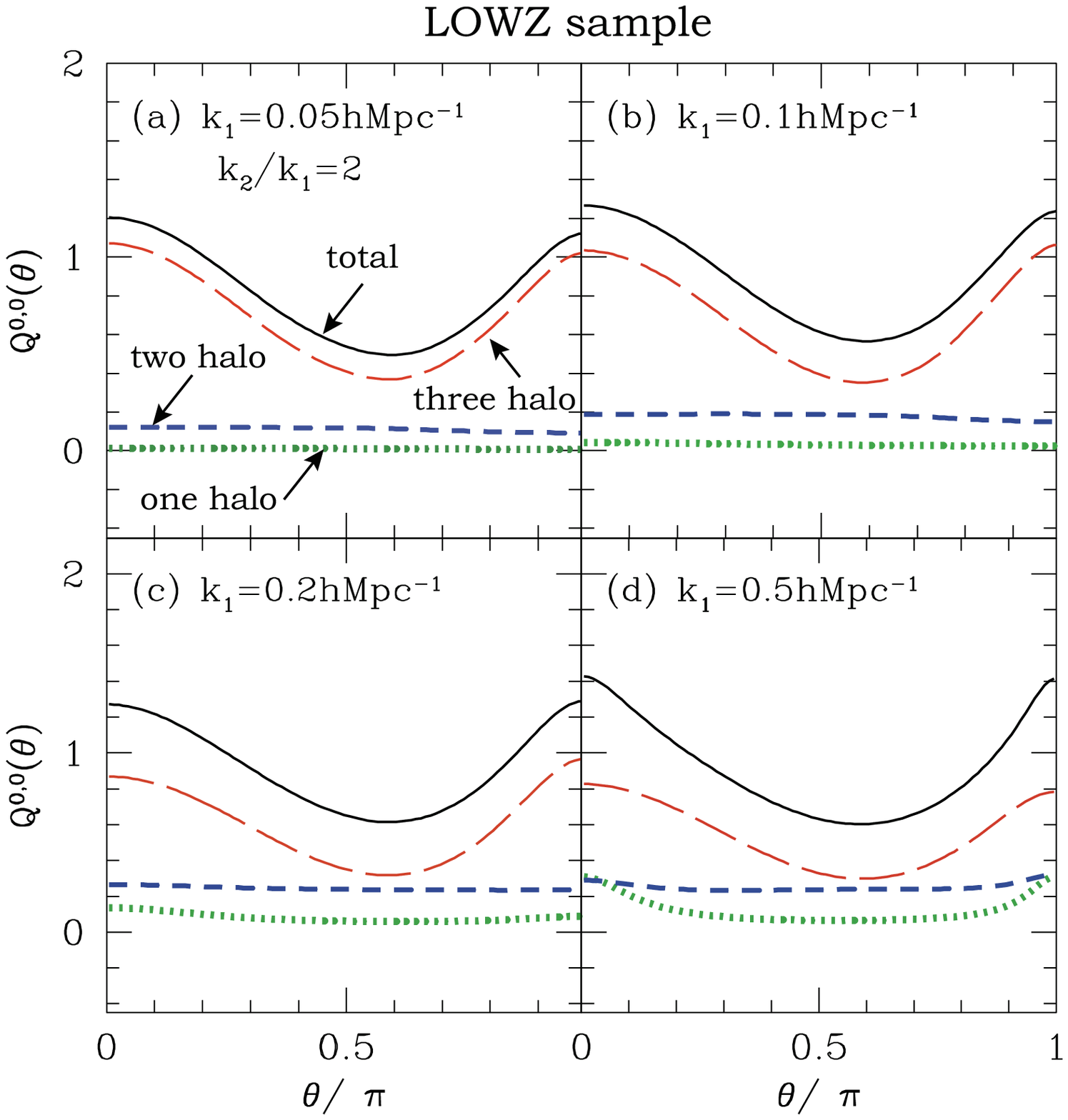}
\caption{Left figure shows $Q^{0,0}(\theta)$ with the LRG sample by fixing (a) $k_1=0.05$, 
(b) $k_1=0.1$, (c) $k_1=0.2$, and (d) $k_1=0.5$ in the unit $h$/Mpc, 
and $k_2/k_1=2$. In each panel the (green) dotted curve is the one-halo term contribution, 
the (blue) short-dashed curve is the two-halo term contribution, the (red) long-dashed curve is
the three-halo term contribution, and the (black) solid curve is the total combination. 
The right figure shows the same as the left figure but for the LOWZ sample. }
 \label{fig:bs9_Q0}
\end{center}
\end{figure}
\begin{figure}[t]
\begin{center}
\includegraphics[width=79mm]{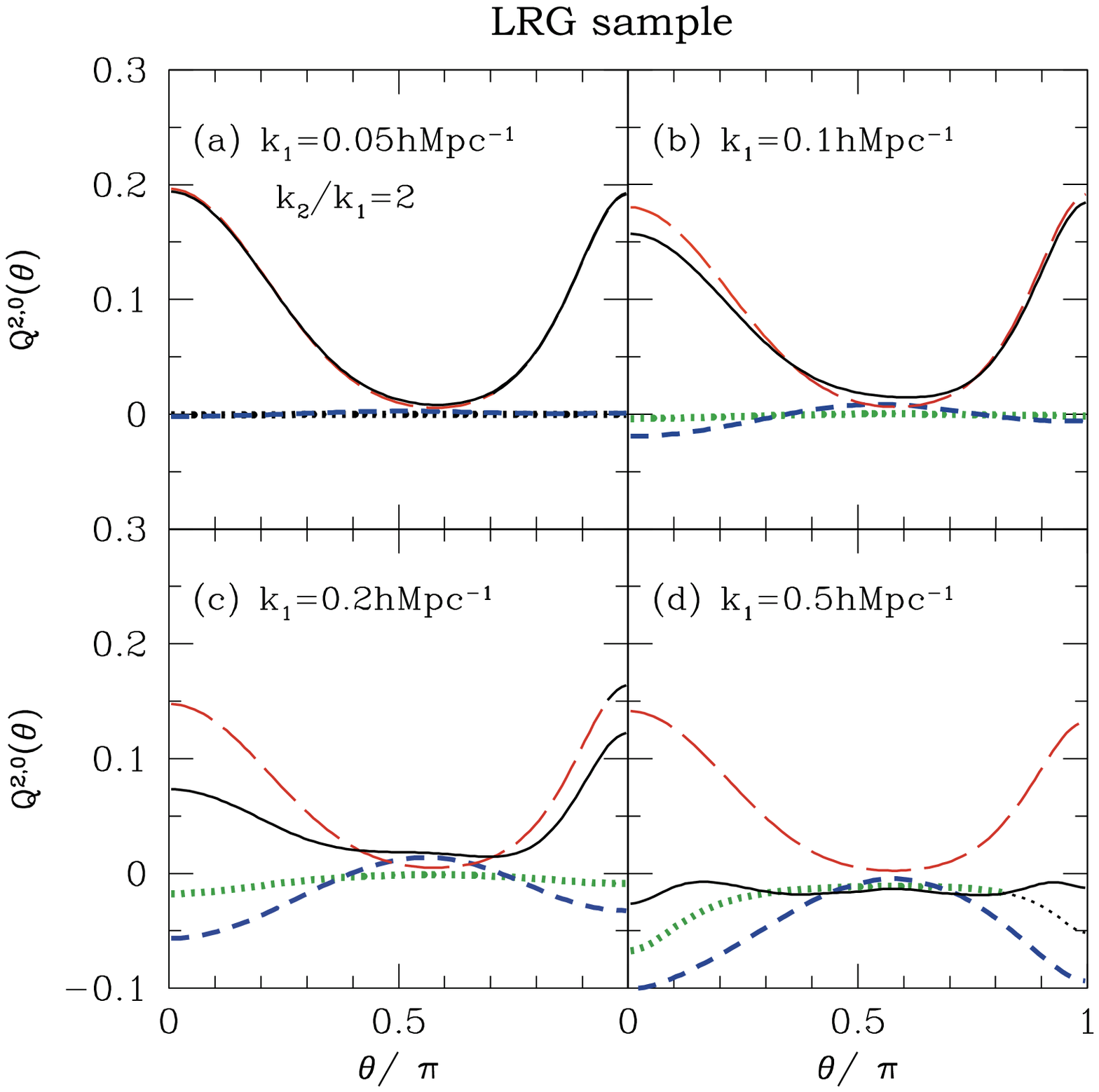}
\hspace{0.9cm}
\includegraphics[width=79mm]{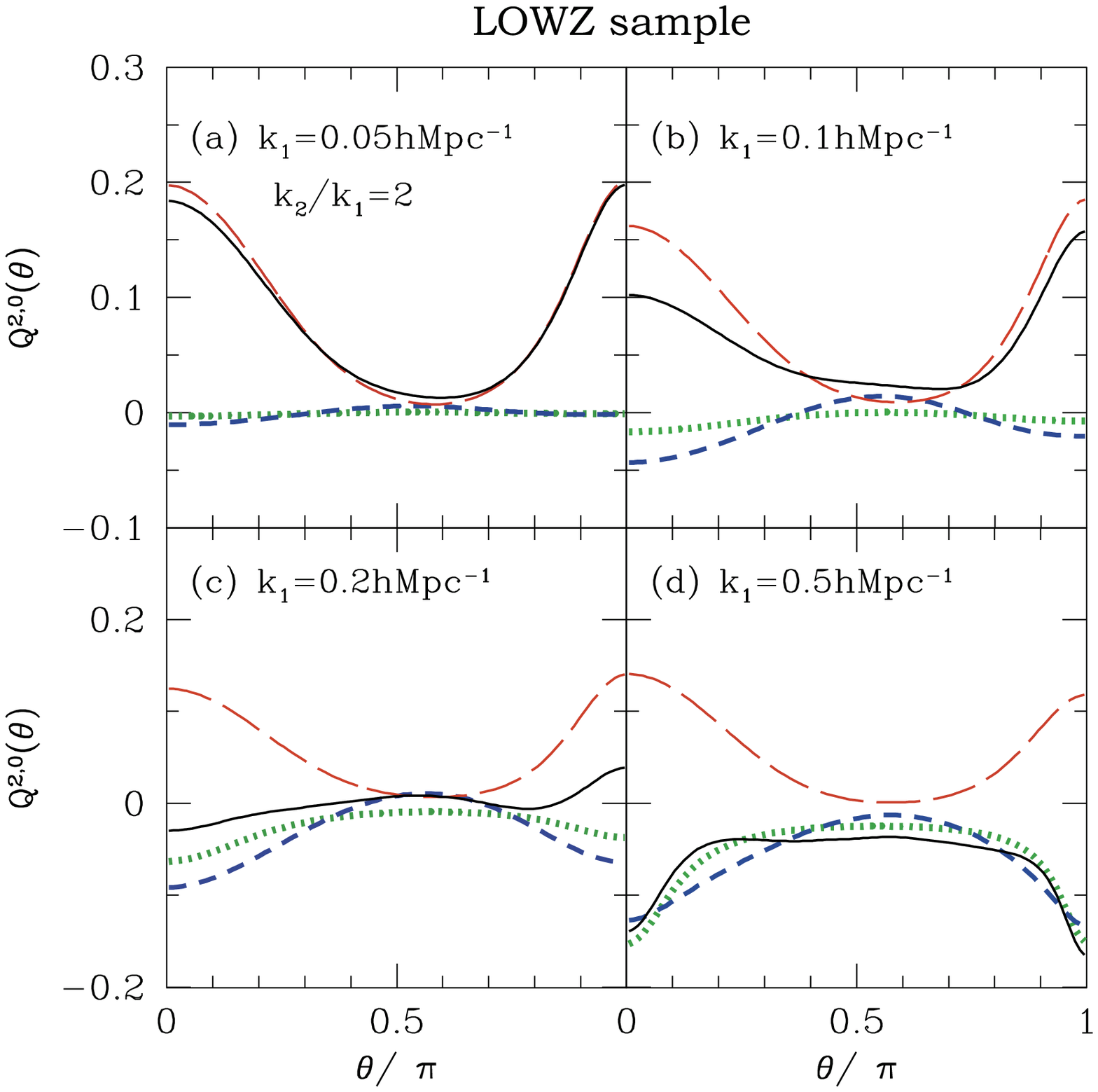}
\caption{Same as Fig.~\ref{fig:bs9_Q0}, but for $Q^{2,0}(\theta)$. 
}
  \label{fig:bs9_Q2}
\end{center}
\vspace{1cm}
\begin{center}
\includegraphics[width=79mm]{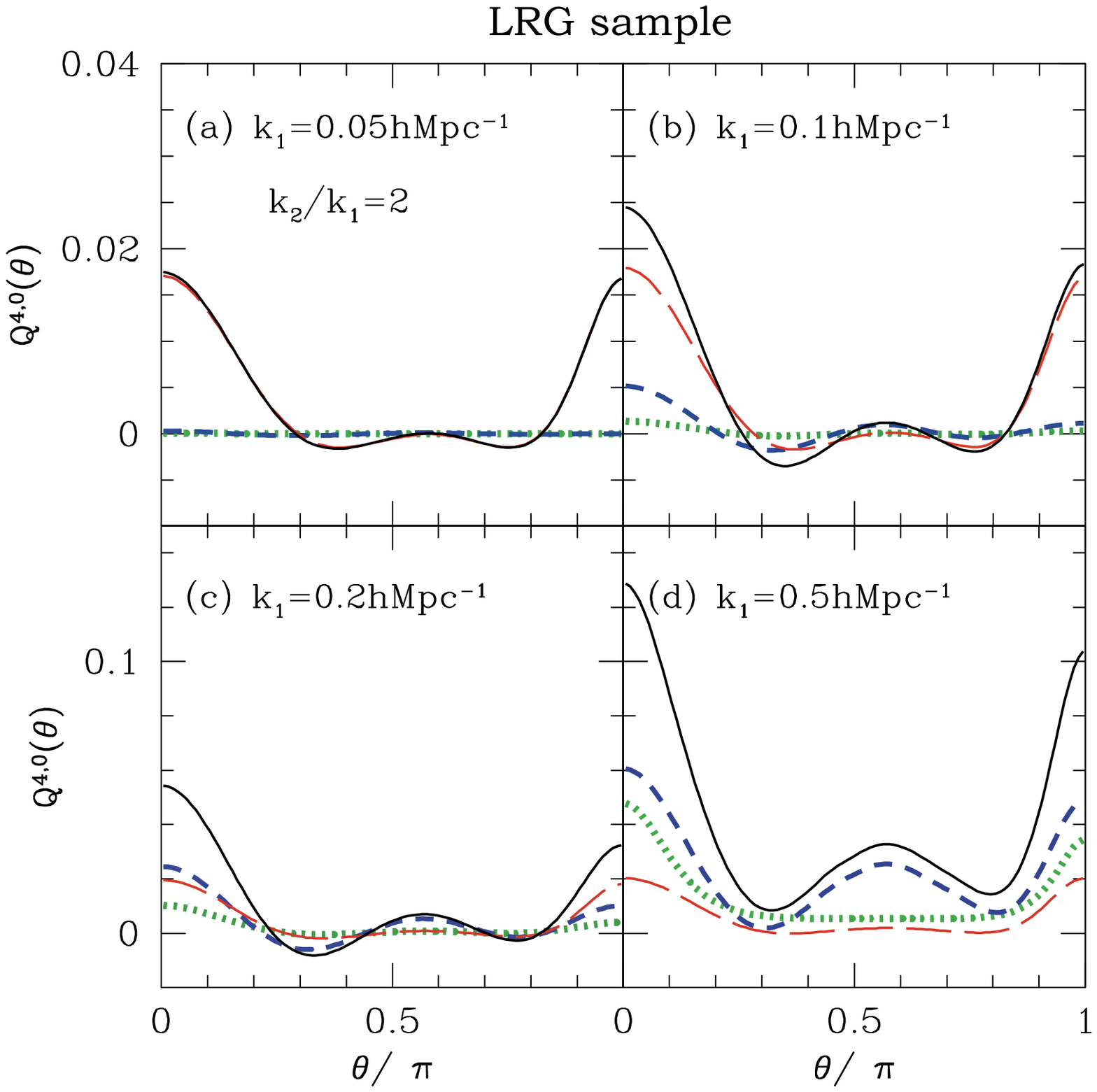}
\hspace{0.9cm}
\includegraphics[width=79mm]{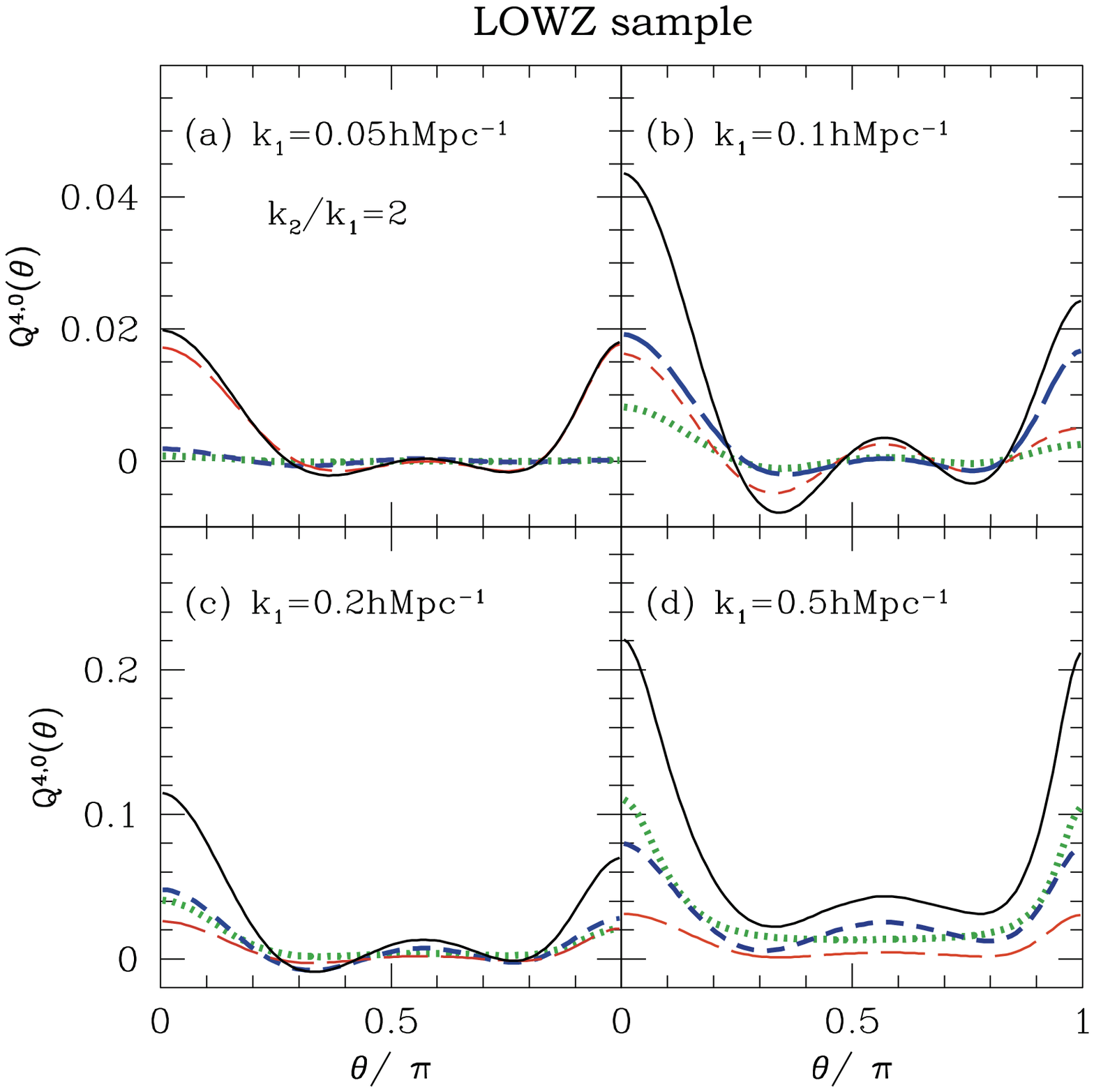}
\caption{Same as Fig.~\ref{fig:bs9_Q0}, but for $Q^{4,0}(\theta)$. }
  \label{fig:bs9_Q4}
\end{center}
\end{figure}

We introduce the multipole of the bispectrum \cite{scoccimarro1999,scoccimarro2015}, 
\begin{eqnarray}
B^{\ell,0}(k_1,k_2,\thetaot)&=&{1\over 4\pi}\int_0^{2\pi} d\phi \int_{-1}^{+1}
d\cos\omega B_g(t,k_1,k_2,\thetaot,\omega,\phi){\cal L}_{\ell}(\cos\omega),
\label{Bsell0}
\end{eqnarray}
where ${\cal L}_\ell(\mu)$ is the Legendre polynomial. Then, we define
the multipoles of the reduced bispectrum as
\begin{eqnarray}
Q^{\ell,0}(k_1,k_2,\thetaot)
={B^{\ell,0}(k_1,k_2,\thetaot) \over P^0(t,k_1)P^0(t,k_2)+P^0(t,k_2)P^0(t,k_3)+P^0(t,k_3)P^0(t,k_1)},
\end{eqnarray}
where $P^0(t,k_i)$ is the monopole spectrum of the galaxy power spectrum $P_g(t,\bm k_i)$,
which is defined by Eq.~(\ref{Pgstk}). 
Because $B_g(t,k_1,k_2,\theta,\omega,\phi)$ consists of the one-halo term, 
the two-halo term, and the three-halo term, $B^{\ell,0}(k_1,k_2,\theta)$ 
and $Q^{\ell,0}(k_1,k_2,\thetaot)$ are written as the sum of the corresponding 
three components,
\begin{eqnarray}
B^{\ell,0}(k_1,k_2,\theta)=
B_{1h}^{\ell,0}(k_1,k_2,\theta)
+B_{2h}^{\ell,0}(k_1,k_2,\theta)
+B_{3h}^{\ell,0}(k_1,k_2,\theta)
\label{BBBB}
\end{eqnarray}
and 
\begin{eqnarray}
Q^{\ell,0}(k_1,k_2,\theta)=
Q_{1h}^{\ell,0}(k_1,k_2,\theta)
+Q_{2h}^{\ell,0}(k_1,k_2,\theta)
+Q_{3h}^{\ell,0}(k_1,k_2,\theta).
\end{eqnarray}


\subsection{Behaviors of the multipoles of the bispectrum}

\subsubsection{Monopole of the bispectrum $Q^{0,0}(k_1,k_2,\thetaot)$}
Figure \ref{fig:bs9_Q0} shows the monopole $Q^{0,0}(k_1,k_2,\thetaot)$ 
as a function of $\theta(=\theta_{12})$, where $k_1$ and $k_2(=2k_1)$ 
are fixed as shown in each panel.
Here the left panels adopt the HOD parameters of the SDSS-II LRG sample, 
while the right panels adopt those of the SDSS-III BOSS LOWZ sample.
$Q^{0,0}_{1h}(k_1,k_2,\theta)$ does not make a significant 
contribution to $Q^{0,0}(k_1,k_2,\theta)$ at scales of $k<0.1~h$/Mpc, 
but it makes non-negligible contribution at scales of $k>0.1~h$/Mpc. 
$Q^{0,0}_{2h}(k_1,k_2,\theta)$ makes significant contribution to 
$Q^{0,0}(k_1,k_2,\theta)$, which is almost constant as a function of $\theta$ 
for $k<0.5~h$/Mpc.

\subsubsection{Higher multipoles of the bispectrum $Q^{2,0}(k_1,k_2,\thetaot)$ and $Q^{4,0}(k_1,k_2,\thetaot)$}
 
Figures~\ref{fig:bs9_Q2} and \ref{fig:bs9_Q4} 
show the same as Fig.~\ref{fig:bs9_Q0} but for $Q^{2,0}(k_1,k_2,\thetaot)$ and $Q^{4,0}(k_1,k_2,\thetaot)$, respectively. 
One can recognize the following features from these figures: 
For the higher multipoles $Q^{\ell,0}(k_1,k_2,\theta)$ with $\ell=2$ and $4$, 
the one-halo term and the two-halo term make a significant contribution
for $k\simgt0.1~h$/Mpc. 
The contribution from the one-halo term and the two-halo term to the multipole bispectrum 
is more significant in the LOWZ sample than that in the LRG sample. 

\begin{figure}[b]
\begin{center}
\vspace{1.0cm}
\includegraphics[width=140mm]{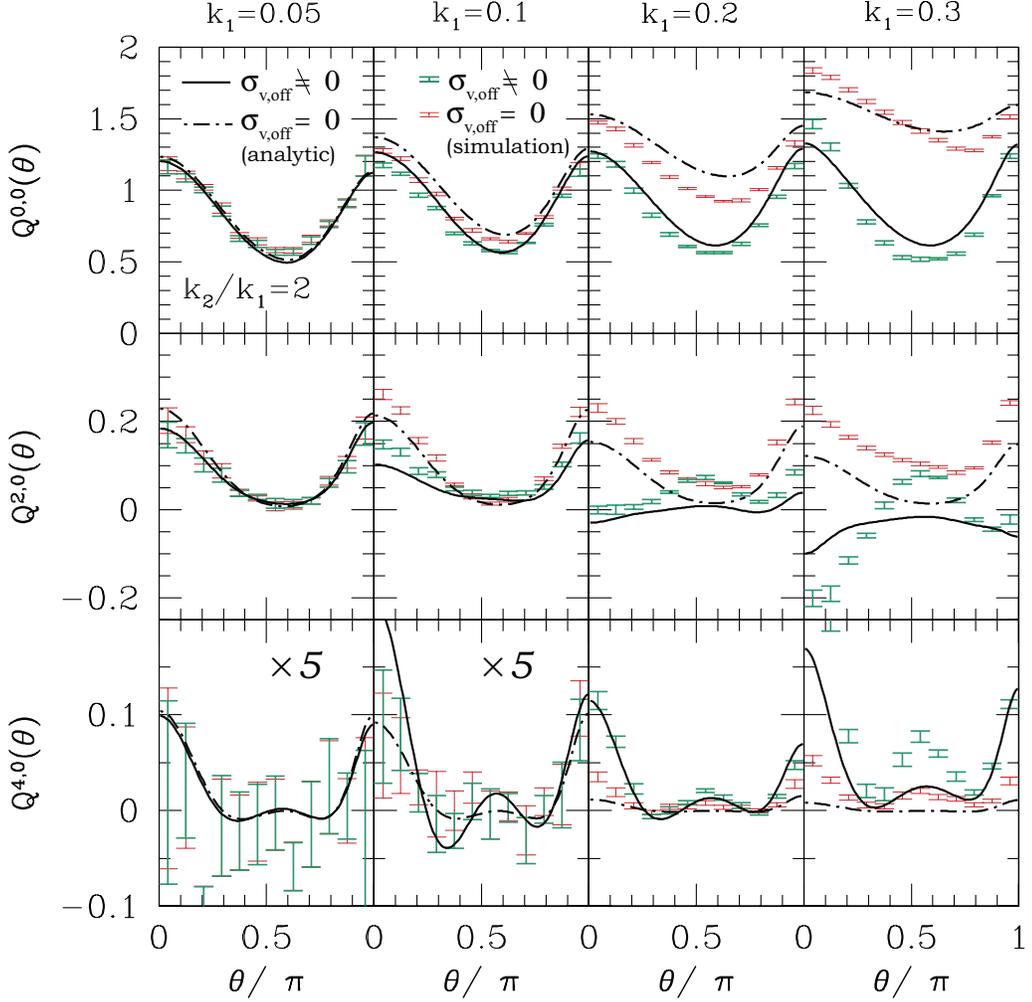}
\caption{Comparison of $Q^{\ell,0}(\theta)$ between our theoretical model and numerical simulations 
adopting the HOD of the LOWZ sample. The top panels, the middle panels, and the bottom panels show 
$Q^{0,0}$, $Q^{2,0}$ and $Q^{4,0}$ as functions of $\theta$. The wave number $k_1/ [h{\rm Mpc}^{-1}]$ 
is fixed as $0.05$, $0.1$, $0.2$ and $0.3$, from left to right, while the ratio of $k_1$ to $k_2$ is 
fixed equally as $k_2/k_1=2$. In each panel, the solid curve is our theoretical model, while 
the dash--dotted curve is the same except for setting $\sigma_{v,{\rm off}}=0$. The meaning of $\sigma_{v,{\rm off}}=0$
is that the random velocity of the satellite galaxy is neglected, thus the FoG effect 
is neglected. The data with error bars are the results from mock catalogs from numerical simulations. 
The thick (green) data points and thin (red) data points respectively correspond to the cases in the presence and absence of 
$\sigma_{v,{\rm off}}$, corresponding to the solid curve and the dash--dotted curve.
All values in the bottom left two panels are multiplied by a factor of $5$.}
  \label{fig:bis_lowz}
\end{center}
\end{figure}

\subsubsection{Comparison with the results of mock catalogs}

For comparison with our analytic model, we construct simulated samples
assuming the HOD of the SDSS-III BOSS LOWZ sample. We run 10
realizations of N-body simulations at the side length of $1h^{-1}$Gpc
with the number of mass particles set as 800$^3$ (mass for each
particle set as $1.3\times 10^{11}h^{-1}M_\odot$) using Gadget-2 code
\citep{Springel05}. The softening length is set to be $50h^{-1}$kpc.
The initial mass distribution is Gaussian starting
from $z=49$ generated by the 2LPT code of \citep{Crocce06}.  We
  adopt a flat $\Lambda$CDM cosmology with the following parameters:
  $\Omega_b=0.046$, $\Omega_m=0.273$, $n_s=0.963$, $h=0.704$,
  $\tau=0.089$, and $\sigma_8=0.809$.  The halo is identified with
the friends-of-friends algorithm with a linking length of 0.2. The
minimum number of mass particles is 10, corresponding to the mass of
$1.3\times 10^{12}h^{-1}M_\odot$. When comparing with our theoretical
model, we add this cut of minimum mass in the integration of mass. The
central and satellite galaxies are assigned to each halo to follow the
HOD of BOSS LOWZ sample. The position and velocity of each central
galaxy are given as the arithmetic mean of all particles in the
halo. The position and velocity of satellites are defined as those of
randomly-selected mass particles.  We confirmed that the mass
resolution of our simulation is sufficient for the following
comparison with our theoretical model.  The data points with error
bars in each panel of Fig.~\ref{fig:bis_lowz} are the results of the
numerical simulations.  The error bars represent 1-sigma dispersion of
$10$ simulation results divided by $\sqrt{10}$, which roughly
corresponds to the sample variance for $10({\rm Gpc}/h)^3$ volume
data.  The binning widths of $k_1, k_2$ and $\theta$ are set to be
$0.01h/$Mpc, $0.02h/$Mpc and $\pi/12$, respectively.

Figure \ref{fig:bis_lowz} shows a comparison between our analytic model of the 
multipoles of the reduced bispectrum and results of numerical simulations 
at various wave numbers of $k_1(=k_2/2)$ (see the caption of Fig.~\ref{fig:bis_lowz}). Here we adopt the HOD of the SDSS-III BOSS LOWZ sample.  
In each panel of this figure, the solid curve is the analytic model, while
the data points with the thick (green) error bars are from the numerical simulations. 
The dash-dotted curve is the analytic model prediction but with settings such 
that the random velocity of the satellite galaxies is zero, i.e., $\sigma_{v,{\rm off}}=0$. 
The data points with the thin (red) error bars are the results of the numerical simulation 
assuming that satellite galaxies have the same velocities as those of the central galaxy 
in the halos, corresponding to the theoretical curve with setting $\sigma_{v,{\rm off}}=0$.
Figure \ref{fig:bis_lowz}  shows that our theoretical model well explains the characteristic 
behavior of the bispectrum from the numerical simulations, though some 
differences arise for the cases with larger wavenumbers at a quantitative level. 
However, the behaviors of the simulations are reproduced at a qualitative level. 
%

\section{Approximate formula} 
We have found that the characteristic behaviors of the multipoles of the bispectrum can be 
explained by our analytic model. Further, we derive analytic approximate formulas for 
the multipoles of the bispectrum, which will be useful to understand the physical 
properties and the origin of the characteristic behaviors of the multipoles of 
the bispectrum. 

According to the case of the multipole power spectrum \cite{HikageYamamoto},
we assume the following approximate formulas for the one-halo term, 
the two-halo term, and the three-halo term,
\begin{eqnarray}
B_{g,1h}(t,{\bm k}_1,{\bm k}_2,{\bm k}_3)&\simeq&{f_s^2\over \bar n^2}
\left(\widetilde u({\bm k}_1)\widetilde u({\bm k}_2)+\widetilde u({\bm k}_2)\widetilde u({\bm k}_3)+\widetilde u({\bm k}_3)\widetilde u({\bm k}_1)\right),
\\
B_{g,2h}(t,{\bm k}_1,{\bm k}_2,{\bm k}_3)
&\simeq&{f_s\over \bar n}
\left(\widetilde u({\bm k}_1)+\widetilde u({\bm k}_2)\right)(\bar b+\mu_3^2f)^2P_m^{\NL}(k_3)
+2~{\rm cyclic~terms},
\\
B_{g,3h}(t,{\bm k}_1,{\bm k}_2,{\bm k}_3)&\simeq&
2P_m^{\NL}(t,k_1)P_m^{\NL}(t,k_2)(\bar b+\mu_1^2f)(\bar b+\mu_2^2f)
\Bigl[\bar bF_2({\bm k}_1,{\bm k}_2)+f \mu_{3}^2G_2({\bm k}_1,{\bm k}_2) +{\bar b_2\over 2}
\nonumber
\\
&&-{1\over 2}f\mu_3k_3
\Bigl({\mu_1\over k_1}(\bar b+f\mu_2^2)+{\mu_2\over k_2}(\bar b+f\mu_1^2)\Bigr)\Bigr]
+2~{\rm cyclic~terms},
\end{eqnarray}
where we use the approximate formula
\begin{eqnarray}
&&\widetilde u({\bm k_i})\simeq 
\exp\left[-\frac{\overline\sigma_{v,{\rm off}}^2k_i^2\mu^2}
{2a^2H^2(z)}\right]=\exp\left[-{\lambda_{v,{\rm off}}^2k_i^2\mu^2}\right]
\end{eqnarray}
for $i=1,2$ and $3$, $\bar b$, and $\overline \sigma_{v,{\rm off}}$ are 
averaged values of the bias and the random velocity of satellite galaxies
over the halo mass, and $f_s$ is the satellite fraction.
Here we introduce the characteristic length scale, associated with the 
random motions by 
\begin{eqnarray}
\lambda_{v{\rm ,off}}^2=\frac{\overline\sigma_{v,{\rm off}}^2}
{2a^2H^2(z)},
\end{eqnarray}
and we include $\bar b_2$ as an averaged value of the bias parameter $b_2$. 

\subsection{One-halo term}
Following the definition of (\ref{Bsell0}) and (\ref{BBBB}), 
by expanding the formulas by the power of $\lambda_{v,{\rm off}}^2$, 
we have the following approximate formulas for the contribution from the one-halo term 
\begin{eqnarray}
B_{1h}^{0,0}(k_1,k_2,\theta)
&\simeq&{f_s^2\over \bar n^2}\left[3-{4\over 3}\left(k_1^2+k_2^2+k_1k_2\cos\thetaot\right)
{\lambda_{v,{\rm off}}^2}\right],
\label{B1h00}
\\
B_{1h}^{2,0}(k_1,k_2,\theta)
&\simeq&-{2 \over 15}{f_s^2\over \bar n^2}\left(
4k_1^2+k_2^2(1+3\cos2\thetaot)+4k_1k_2\cos\thetaot\right){\lambda_{v,{\rm off}}^2},
\label{B1h20}
\\
B_{1h}^{4,0}(k_1,k_2,\theta)&=&{\cal O}\left({f_s^2\over \bar n^2}\left({k^2 \lambda_{v,{\rm off}}^2}\right)^2\right).
\end{eqnarray}
Thus, the contribution from the one-halo term to the multipole bispectrum is on the order of $f_s^2/\bar n^2$. 
The dominant term in the monopole $B_{1h}^{0,0}$  is constant $3f_s^2/\bar n^2$ for 
$k^2\lambda_{v,{\rm off}}^2\ll1$ from (\ref{B1h00}), 
however, $k$ and $\theta$ dependence arises for $k^2\lambda_{v,{\rm off}}^2
\simgt1$. The order of higher multiple bispectrum $B^{\ell,0}_{{1h}}$ is roughly ${\cal O}((f_s^2/\bar n^2)
(k\lambda_{v,{\rm off}})^\ell)$ for $\ell=2$ and $4$.

Thus, one can roughly write the approximate formulas for
term contribution to the multiple bispectrum as 
\begin{eqnarray}
&&B_{1h}^{0,0}(k_1,k_2,\theta)\sim \left({f_s\over \bar n}\right)^2\left(3-{\cal O}(k^2\lambda_{v,{\rm off}}^2)\right),
\label{b1haloterm00}
\\
&&B_{1h}^{2,0}(k_1,k_2,\theta)\sim \left({f_s\over \bar n}\right)^2{\cal O}\left(k^2\lambda_{v,{\rm off}}^2\right),
\label{b1haloterm20}
\\
&&B_{1h}^{4,0}(k_1,k_2,\theta)\sim \left({f_s\over \bar n}\right)^2{\cal O}\left(k^4\lambda_{v,{\rm off}}^4\right).
\label{b1haloterm40}
\end{eqnarray}

\subsection{Two-halo term}
Similarly, one can derive the contribution from the two halo term to the monopole bispectrum
and the quadrupole bispectrum as 
\begin{eqnarray}
&&B_{2h}^{0,0}(k_1,k_2,\theta)
\simeq{f_s\over \bar n}\Bigl[
\Bigl(2\bar b^2+{4bf\over 3}+{2f^2\over 5}\Bigr)(P_m(k_1)+P_m(k_2)+P_m(k_3))
\nonumber\\
&&~~~~
+{\lambda_{v,{\rm off}}^2\over 105}\Bigl\{
{P_m(k_1)}\Bigl(-(35b^2+42bf+15f^2)k_1^2
-2(35\bar b^2+28\bar bf+9f^2)k_2^2
-2(35\bar b^2+42\bar bf+15f^2)k_1k_2\cos\thetaot
\nonumber\\
&&~~~~
-4f(7\bar b+3f)k_2^2\cos2\thetaot\Bigr)
+{P_m(k_2)}\Bigl(-2(35\bar b^2+28\bar bf+9f^2)k_1^2
-(35\bar b^2+42\bar bf+15f^2)k_2^2
-2(35\bar b^2+42\bar bf
\nonumber\\
&&~~~~
+15f^2)k_1k_2\cos\thetaot
-4f(7\bar b+3f)k_1^2\cos2\thetaot\Bigr)
-{P_m(k_3)\over k_1^2+k_2^2+2k_1k_2\cos\thetaot}
\Bigl(35\bar b^2(k_1^2+k_2^2)^2
+14\bar bf(3k_1^4+4k_1^2k_2^2
\nonumber\\
&&~~~~
+3k_2^4)
+3f^2(5k_1^4+6k_1^2k_2^2+5k_2^4)
+2k_1k_2\cos\thetaot(35\bar b^2+42\bar bf+15f^2)(k_1^2+k_2^2)+2f(7\bar b+3f)k_1k_2\cos2\thetaot\Bigr)
\Bigr\}\Bigr],
\nonumber\\
\label{B2h00}
\end{eqnarray}
\begin{eqnarray}
&&B_{2h}^{2,0}(k_1,k_2,\theta)
\simeq{f_s\over 210\bar n}\Bigl[{16f(7\bar b+3f)}P_m(k_1)+{4f(7\bar b+3f)}
(1+3\cos2\thetaot)P_m(k_2)
\nonumber\\
&&~~~~
+{4f(7\bar b+3f)(4k_1^2+k_2^2+8k_1k_2\cos\thetaot+3k_2^2\cos2\thetaot)\over 
k_1^2+k_2^2+2k_1k_2\cos\thetaot}P_m(k_3)
+{\lambda_{v,{\rm off}}^2 }\Bigl\{{2P_m(k_1)}\Bigl(-2(\bar b+f)(7\bar b+5f)k_1^2
\nonumber\\
&&~~~~
-(7\bar b^2+26\bar bf+11f^2)k_2^2
-4(\bar b+f)(7\bar b+5f)k_1k_2\cos\thetaot-(21\bar b^2+22\bar bf+9f)k_2^2\cos2\thetaot\Bigr)
\nonumber\\
&&~~~~+{P_m(k_2)}\Bigl(-2(28\bar b^2+26\bar bf+9f^2)k_1^2-(\bar b+f)(7\bar b+5f)k_2^2
-2(28\bar b^2+39\bar bf+15f^2)k_1k_2\cos\thetaot
\nonumber\\
&&~~~~
-(4f(11\bar b+5f)k_1^2+3(\bar b+f)(7\bar b+5f)k_2^2)\cos2\thetaot-2fk_1((9\bar b+5f)k_2\cos3\thetaot
+fk_1\cos4\thetaot)\Bigr)
\nonumber\\
&&~~~~
-{P_m(k_3)\over (k_1^2+k_2^2+2k_1k_2\cos\thetaot)^2}\Bigl(
4(\bar b+f)(7\bar b+5f)k_1^6 + 7(\bar b+f)(17\bar b +11f)k_1^4 k_2^2 
+ (77 \bar b^2 + 142 \bar b f 
\nonumber\\
&&~~~~
+     63 f^2) k_1^2 k_2^4 
+ (\bar b + f) (7 \bar b + 5 f) k_2^6 +  2 k_1 k_2 
\Bigl\{8 (\bar b + f) (7 \bar b + 5 f) k_1^4 
+ (91 \bar b^2 + 161 \bar b f + 66 f^2) k_1^2 k_2^2 
\nonumber\\
&&~~~~
+ (35 \bar b^2 + 69 \bar b f + 30 f^2) k_2^4\Bigr\} \cos\thetaot 
+ k_2^2 \Bigl\{7 (\bar b + f) (11 \bar b + 9 f) k_1^4 
+ 2 (49 \bar b^2 + 88 \bar b f + 35 f^2) k_1^2 k_2^2 
\nonumber\\
&&~~~~
+ 3 (\bar b + f) (7 \bar b + 5 f) k_2^4\Bigr\} \cos2\thetaot 
+ 2k_1 k_2^3 \Bigl\{(21 \bar b^2 + 31 \bar b f + 14 f^2) k_1^2 
+ (21 \bar b^2 + 27 \bar b f + 10 f^2) k_2^2\Bigr\} \cos 3 \thetaot 
\nonumber\\
&&~~~~
+ (21 \bar b^2 + 18 \bar b f + 7 f^2) k_1^2 k_2^4 \cos4 \thetaot\Bigr)\Bigr\}\Bigr].
\label{B2h20}
\end{eqnarray}
Thus, the contribution of the two-halo term to the multipole bispectrum is in proportion to $f_s/\bar n$. 
The dominant term in the monopole bispectrum from the two-halo term is roughly on 
the order of $2(f_s/\bar n)(P_{g}^{0}(k_1)+P_{g}^{0}(k_2)
+P_{g}^0(k_3))$ for $k^2\lambda_{v,{\rm off}}^2\ll1$, where $P_g^0(k)$ is the monopole galaxy power spectrum 
in linear theory, defined by 
\begin{eqnarray}
P_{g}^{\ell}(k)={1\over 2}\int_{-1}^{+1}d\mu \left(\bar b+f\mu^2\right)^2P_m(k) {\cal L}_{\ell}(\mu)
\end{eqnarray}
with $\ell=0$. Explicitly, we write
\begin{eqnarray}
&&P_{g}^{0}(k)=\left(\bar b^2+{2\bar bf\over 3}+{f^2\over 5}\right)P_m(k),
\\
&&P_{g}^{2}(k)=\left({4\bar bf\over 15}+{4f^2\over 35}\right)P_m(k),
\end{eqnarray}
for $\ell=0$ and $2$. 
Similarly, the dominant terms of the quadrupole bispectrum $B_{2h}^{2,0}$ 
from the two-halo term contribution are $(f_s/\bar n)P^2_g(k_i)$ for $k^2\lambda_{v,{\rm off}}^2\ll1$. 
However, other contributions depending on $k$ and $\theta$ significantly emerge
for $k^2\lambda_{v,{\rm off}}^2\simgt1$. These properties are common to $B_{2h}^{4,0}(k_1,k_2,\theta)$. 
For the higher multipoles, $B_{2h}^{\ell,0}(k_1,k_2,\theta)$ with $\ell\geq 2$, 
the contributions of the terms in proportion to $\lambda_{\rm off}^{2\ell}$ are important.

In summary, the two-halo term contribution to the multipole bispectrum is more 
important than that of the one-halo term, and can be roughly expressed as
\begin{eqnarray}
&&B_{2h}^{\ell,0}(k_1,k_2,\theta)\sim 
\left({2f_s\over \bar n}\right)\left({\cal O}(P^\ell_g(k_1)+P^\ell_g(k_2)+P^\ell_g(k_3))
-{\cal O}(P_g(k) k^2\lambda^2_{v,{\rm off}})\right)
\label{jsofdjoas}
\end{eqnarray}
for $\ell=0$ and $2$.

\subsection{Three-halo term}
For an analytic description, we complete this section by presenting the 
analytic formula for the multipole bispectrum from the three-halo terms,
which can be written as
\begin{eqnarray}
B_{3h}^{\ell,0}(k_1,k_2,\theta)=C^{\ell}_{12}P_m(k_1)P_m(k_2)+C^{\ell}_{23}P_m(k_2)P_m(k_3)+C^{\ell}_{31}P_m(k_3)P_m(k_1),
\end{eqnarray}
where $C^\ell_{12}$, $C^\ell_{23}$ and $C^\ell_{31}$ are listed in the Appendix.

The mathematical formulas of this section and those of the appendix are derived 
using \emph{Mathematica}. The source \emph{Mathematica} programs for the derivation 
are included in the source file uploaded at the site arXiv.org (Ref.~\cite{YamamotoNanHikage}).

\subsection{Discussions}

We find that $B_{1h}^{0,0}(k_1,k_2,\theta)$ and $B_{2h}^{0,0}(k_1,k_2,\theta)$ are symmetric 
with respect to the exchange of wavenumbers between $k_1$ and $k_2$. However, the higher multipoles
$B_{1h}^{\ell,0}(k_1,k_2,\theta)$ and $B_{2h}^{\ell,0}(k_1,k_2,\theta)$ with $\ell\geq2$ 
do not show the symmetric property.

In the above expressions, (\ref{B1h00})  and (\ref{B2h00}), the contribution to the monopole bispectrum 
from the one-halo term and from the two-halo term, respectively, 
the dominant terms in the limit $k^2\lambda_{v,{\rm off}}^2\ll1$ are positive and almost 
constant as functions of $\theta$. 
On the other hand, for $k^2\lambda_{v,{\rm off}}^2\simgt1$, they depend on $k$ and $\theta$.
These properties explain the characteristic behaviors of their contribution to the 
monopole bispectrum in Fig.~\ref{fig:bs9_Q0}. 

From the expression (\ref{B1h20}), the contribution to the quadrupole bispectrum 
from the one-halo term,  
we read that the one-halo term contribution to the quadrupole bispectrum is negative. 
We also find that the expression (\ref{B2h20}), the contribution to the quadrupole bispectrum 
from the two-halo term, is positive in the limit $k^2\lambda_{v,{\rm off}}^2\ll1$, 
but the terms in proportion to $\lambda_{v,{\rm off}}^2$ are negative.
This can be easily checked for $\theta=0$.   
The latter terms explain the behaviors of the quadrupole bispectrum for 
$k\simgt 0.1$ in Fig.~\ref{fig:bs9_Q2}. 
Thus, the FoG effect is important for the higher multiple bispectrum.

\section{Summary and Conclusions} 
%

In summary, we have developed an analytic model of the redshift space bispectrum 
based on the halo approach with the HOD with central and satellite galaxies. 
We have demonstrated characteristic behaviors of the multipoles of the 
bispectrum depending on the HOD parameters of galaxy samples. In particular, 
the contribution from the two-halo term to the multipole bispectrum is important 
at the scales $k\simgt 0.1h$/Mpc. 
The one-halo term makes a non-negligible contribution to $Q^{\ell,0}$ at the scales $k\simgt 0.1h$/Mpc. 
The influences from the one-halo term and the two-halo term are more significant
for the higher multipole bispectrum $Q^{\ell,0}$ with $\ell\geq2$. 

Based on our analytic approach, we have derived the approximate formulas for the 
multipoles of the bispectrum. Summarizing the results in section~4, we can write the one-halo 
term contribution to the multiple bispectrum, Eqs. (\ref{b1haloterm00}), (\ref{b1haloterm20}), 
and (\ref{b1haloterm40}).
In the SDSS-II LRG sample and SDSS-III BOSS LOWZ sample, the satellite fraction is small, 
$5$\% and $11$\%, 
respectively, and the number of halos containing more than three galaxies is small.
Therefore, the one-halo term contribution is smaller than the two-halo term 
contribution. 
In general, the two-halo term contribution to the multipole bispectrum is more 
important, and can be roughly expressed by Eq.~(\ref{jsofdjoas}) for the 
monopole and the quadrupole. 
The two-halo term contribution to the monopole bispectrum, $B_{2h}^{0,0}$, 
makes a positive and constant contribution, but the two-halo term contribution 
to the quadrupole bispectrum, $B_{2h}^{2,0}$, take a significant negative value, 
where the terms in proportion to $\lambda_{v,{\rm off}}^2$ are important.
For the hexadecapole bispectrum, the terms in proportion to $\lambda_{v,{\rm off}}^4$ 
make a significant contribution. 
The one-halo term and the two-halo term make quite significant contributions to the quadrupole 
bispectrum on the scales $k\simgt 0.1~h$/Mpc. 
These observations are useful to explain the numerical results of our theoretical model 
as well as those from numerical simulations in section 3. 

Thus, we have found that our model well describes the simulated bispectrum, including 
the FoG effect. 
The disagreement between the model expectations and the simulated results, however, 
increases as the scale goes nonlinear. One of the reasons for this disagreement comes 
from the approximation of the halo clustering terms $P_{2h}$ (\ref{twohalom}) and $P_{3h}$ 
(\ref{threehalom}) with linear Kaiser formulas. We expect that the agreement would improve 
by directly using the simulated results of these terms instead of the Kaiser approximation. 
We leave this to work in the near future. 

In our analysis, we adopted the fitting formula for the halo mass function (\ref{massfunctionmodel})
and for the bias model (\ref{biasmodel}).
We may use other formula for the bias, e.g., in Ref. \cite{ShethMoTormen2001}, 
and for the halo mass function e.g., in Refs. \cite{Jenkins,Tinker2008}.
We checked how our results depend on the models the bias and the halo mass function.
Our conclusions of the present paper are not altered qualitatively, however, 
the results depend on the models for the bias and the halo mass function, 
at a quantitative level. The modification is not so significant but it cannot be 
negligible, especially for the higher bispectrum $Q^{\ell,0}$ with $\ell \geq 2$. 
The results suggest that the uncertainties of these properties cause difficulties
in precisely predicting the bispectrum. Conversely, the information of about the 
bias, the mass function, and the galaxy-halo connection are included in the 
bispectrum of galaxies, therefore, the validity of their models might be 
investigated through the observational measurements.

\section*{Acknowledgment}
This work was supported by MEXT/JSPS KAKENHI Grant Number 15H05895 and JP16H03977. 
We thank A. Taruya, I. Hashimoto, and M. Takada for useful comments.

\begin{appendix}
\def\barb{\bar b}
\begin{eqnarray}
&&C^0_{12}={1\over 4410 k_1 k_2 (k_1^2 + k_2^2 + 
   2 k_1 k_2 \cos\thetaot)}\bigl[
3 \bigl\{210 \barb^2 (19 \barb + 7 \barb_2) + 70 \barb (\barb (55 + 21 \barb) + 14 \barb_2) f 
\nonumber\\
&&~~
+ 
    49 (\barb (39 + 46 \barb) + 4 \barb_2) f^2 
+ 3 (131 + 448 \barb) f^3 + 
    308 f^4\bigr\} k_1 k_2 (k_1^2 + k_2^2) + 
 2 \bigl\{21 (35 \barb^3 (3 + f) 
\nonumber\\
&&~~
+ 14 \barb f^2 (4 + 3 f) + 21 \barb^2 f (5 + 3 f) + 
       2 f^3 (6 + 5 f)) k_1^4 
+ \bigl(630 \barb^3 (20 + 7 f) + 
       210 \barb^2 (21 \barb_2 + f (59 + 35 f)) 
\nonumber\\
&&~~
+ 
       42 \barb f (70 \barb_2 + f (161 + 114 f)) + 
       f^2 (735 \barb_2 + 2 f (747 + 574 f))\bigr) k_1^2 k_2^2 
+ 
    21 (35 \barb^3 (3 + f) + 14 \barb f^2 (4 + 3 f) 
\nonumber\\
&&~~
+ 21 \barb^2 f (5 + 3 f) + 
       2 f^3 (6 + 5 f)) k_2^4\bigr\} \cos\theta + 
 2 \bigl\{105 \barb^3 (27 + 7 f) + 21 \barb f^2 (109 + 81 f) 
+ 
    21 \barb^2 f (145 + 91 f)
\nonumber\\
&&~~
 + 
    f^2 (147 \barb_2 + 67 f (9 + 7 f))\bigr\} k_1 k_2 (k_1^2 + k_2^2) \cos 2\theta + 
 14 \bigl\{f^2 (3 \barb (7 + 3 f) + f (9 + 5 f)) k_1^4 
+ (90 \barb^3 + 120 \barb^2 f 
\nonumber\\
&&~~
+ 
       3 (\barb (55 + 28 \barb) + 7 \barb_2) f^2 + 18 (3 + 7 \barb) f^3 + 
       44 f^4) k_1^2 k_2^2 + 
    f^2 (3 \barb (7 + 3 f) + f (9 + 5 f)) k_2^4\bigr\} \cos3\theta 
\nonumber\\
&&~~
+ 
 f^2 (21 \barb (13 + 6 f) + f (135 + 98 f)) k_1 k_2 (k_1^2 + k_2^2) \cos4\theta + 
 2 f^2 (21 \barb + 2 f (9 + 7 f)) k_1^2 k_2^2 \cos5\theta
\bigr]
\\
&&C^0_{23}={1\over 4410 k_2 (k_1^2 + k_2^2 + 
   2 k_1 k_2 \cos\thetaot)^2}k_1^2\bigl[
-3 \bigl\{210 \barb^2 (9 \barb - 7 \barb_2) + 70 \barb (\barb (29 + 7 \barb) - 14 \barb_2) f 
\nonumber\\
&&~~+ 
    49 (\barb (25 + 26 \barb) - 4 \barb_2) f^2 + 9 (31 + 112 \barb) f^3 + 
    252 f^4\bigr\} k_1^4 k_2 - 
 2 \bigl\{630 \barb^2 (\barb - 14 \barb_2) + 840 \barb (\barb - 7 \barb_2) f 
\nonumber\\
&&~~
+ 
    21 (\barb (25 + 28 \barb) - 77 \barb_2) f^2 + 9 (15 + 56 \barb) f^3 + 
    140 f^4\bigr\} k_1^2 k_2^3 + 294 \barb_2 \bigl(15 \barb^2 + 10 \barb f + 3 f^2\bigr) k_2^5 
\nonumber\\
&&~~
- 
 6 k_1 \bigl\{7 \bigl(35 \barb^3 (3 + f) + 14 \barb f^2 (4 + 3 f) + 21 \barb^2 f (5 + 3 f) + 
       2 f^3 (6 + 5 f)\bigr) k_1^4 + \bigl(420 \barb^2 (4 \barb - 7 \barb_2) 
\nonumber\\
&&~~+ 
       70 \barb (\barb (25 + 7 \barb) - 28 \barb_2) f + 
       7 (\barb (145 + 154 \barb) - 77 \barb_2) f^2 + 3 (79 + 266 \barb) f^3 + 
       210 f^4\bigr) k_1^2 k_2^2 
\nonumber\\
&&~~
- 
    196 \barb_2 (15 \barb^2 + 10 \barb f + 3 f^2) k_2^4\bigr\} \cos\theta  
 -6k_1^2k_2 \bigl\{\bigl(70 \barb^3 (18 + 7 f) + 14 \barb f^2 (47 + 30 f) + 
          14 \barb^2 f (85 + 49 f) 
\nonumber\\
&&~~
+ 
          f^2 (-49 \barb_2 + 4 f (39 + 28 f))\bigr) k_1^2 + \bigl(35 \barb^3 (15 + 7 f) + 
          49 \barb f (-20 \barb_2 + f (5 + 3 f)) + 
          35 \barb^2 (-42 \barb_2 + f (13 + 7 f)) 
\nonumber\\
&&~~
+ 
          f^2 (-343 \barb_2 + f (51 + 35 f))\bigr) k_2^2\bigr\} \cos 2\theta 
-     7 k_1 \bigl\{2 f^2 (3 \barb (7 + 3 f) + f (9 + 5 f)) k_1^2 + 
       3 \bigl(10 \barb^2 f (13 + 7 f) 
\nonumber\\
&&~~
+
 10 \barb^3 (15 + 7 f) + 
          \barb f^2 (85 + 42 f) + f^2 (-14 \barb_2 + f (21 + 10 f))\bigr) k_2^2\bigr\} \cos3\theta - 
    3 f^2 k_2 \bigl\{(21 \barb (5 + 2 f) 
\nonumber\\
&&~~
+ f (39 + 14 f)) k_1^2 + 
       6 (7 \barb + 3 f) k_2^2\bigr\} \cos4\theta - 
    9 f^2 (7 \barb + 3 f) k_1 k_2^2 \cos5\theta
\bigr]
\\
&&C^0_{31}={1\over4410 k_1 (k_1^2 + k_2^2 + 
   2 k_1 k_2 \cos\thetaot)^2}k_2^2 \bigl[
294 \barb_2 (15 \barb^2 + 10 \barb f + 3 f^2) k_1^5 - 
 2 \bigl(630 \barb^2 (\barb - 14 \barb_2) + 840 \barb (\barb - 7 \barb_2) f 
\nonumber\\
&&~~
+ 
    21 (\barb (25 + 28 \barb) - 77 \barb_2) f^2 + 9 (15 + 56 \barb) f^3 + 
    140 f^4\bigr) k_1^3 k_2^2 - 
 3 \bigl(210 \barb^2 (9 \barb - 7 \barb_2) + 70 \barb (\barb (29 + 7 \barb) - 14 \barb_2) f 
\nonumber\\
&&~~
+ 
    49 (\barb (25 + 26 \barb) - 4 \barb_2) f^2 + 9 (31 + 112 \barb) f^3 
+ 
    252 f^4\bigr) k_1 k_2^4 - 
 6 k_2 \bigl\{-196 \barb_2 (15 \barb^2 + 10 \barb f + 
       3 f^2) k_1^4 
\nonumber\\
&&~~+ \bigl(420 \barb^2 (4 \barb - 7 \barb_2) + 
       70 \barb (\barb (25 + 7 \barb) - 28 \barb_2) f + 
       7 (\barb (145 + 154 \barb) - 77 \barb_2) f^2 + 3 (79 + 266 \barb) f^3 + 
       210 f^4\bigr) k_1^2 k_2^2 
\nonumber\\
&&~~
+ 
    7 \bigl(35 \barb^3 (3 + f) + 14 \barb f^2 (4 + 3 f) + 21 \barb^2 f (5 + 3 f) + 
       2 f^3 (6 + 5 f)\bigr) k_2^4\bigr\} \cos\theta 
 -6k_2^2  k_1 \bigl\{\bigl(35 \barb^3 (15 + 7 f) 
\nonumber\\
&&~~
+ 49 \barb f (-20 \barb_2 + f (5 + 3 f)) + 
          35 \barb^2 (-42 \barb_2 + f (13 + 7 f)) + 
          f^2 (-343 \barb_2 + f (51 + 35 f))\bigr) k_1^2 + (70 \barb^3 (18 + 7 f) 
\nonumber\\
&&~~
+ 
          14 \barb f^2 (47 + 30 f) 
+ 14 \barb^2 f (85 + 49 f) + 
          f^2 (-49 \barb_2 + 4 f (39 + 28 f))) k_2^2\bigr\} \cos 2\theta 
- 
    7 k_2 \bigl\{3 (10 \barb^2 f (13 + 7 f) 
\nonumber\\
&&~~
+ 10 \barb^3 (15 + 7 f) + 
          \barb f^2 (85 + 42 f) + f^2 (-14 \barb_2 + f (21 + 10 f))) k_1^2 + 
       2 f^2 (3 \barb (7 + 3 f) + f (9 + 5 f)) k_2^2\bigr\} \cos3\theta 
\nonumber\\
&&~~
- 
    3 f^2 k_1 \bigl\{6 (7 \barb + 3 f) k_1^2 + (21 \barb (5 + 2 f) + 
          f (39 + 14 f)) k_2^2\bigr\} \cos4\theta - 
    9 f^2 (7 \barb + 3 f) k_1^2 k_2 \cos5\theta
\bigr]
\end{eqnarray}

\begin{eqnarray}
&&\hspace{-0.5cm}
C^2_{12}={f\over 97020 k_1k_2 (k_1^2 + k_2^2 + 
     2 k_1 k_2 \cos\thetaot)}\bigl[
3 \bigl(9702 \barb^3 + 77 \barb^2 (353 + 299 f) + 
      22 \barb (245 \barb_2 + f (993 + 791 f)) 
\nonumber\\
&&~~
+ 
      f (2002 \barb_2 + f (5357 + 4487 f))\bigr) k_1^3 k_2 + 
   33 \bigl(882 \barb^3 + 35 \barb^2 (61 + 58 f) + 2 \barb (245 \barb_2 + 849 f + 756 f^2) 
\nonumber\\
&&~~
+
       f (182 \barb_2 + f (451 + 392 f))\bigr) k_1 k_2^3 + \bigl\{21 \bigl(616 \barb^3 + 
         33 \barb^2 (49 + 45 f) + 22 \barb f (61 + 53 f) + 
         f^2 (341 + 305 f)\bigr) k_1^4 
\nonumber\\
&&~~
+ \bigl(231 \barb (\barb (809 + 294 \barb) + 
            182 \barb_2) + 231 (\barb (670 + 713 \barb) + 74 \barb_2) f + 
         33 (1233 + 3892 \barb) f^2 + 34069 f^3\bigr) k_1^2 k_2^2 
\nonumber\\
&&~~
+ 
      42 \bigl(308 \barb^3 + 693 \barb^2 (1 + f) + 14 f^2 (11 + 10 f) + 
         44 \barb f (13 + 12 f)\bigr) k_2^4\bigr\} \cos\theta + 
   k_1 k_2 \bigl\{2 \bigl(231 \barb (\barb (146 + 49 \barb) 
\nonumber\\
&&~~
+ 21 \barb_2) + 
         33 (920 \barb + 938 \barb^2 + 77 \barb_2) f + 33 (274 + 819 \barb) f^2 + 
         7756 f^3\bigr) k_1^2 + \bigl(462 \barb (\barb (164 + 49 \barb) + 21 \barb_2) 
\nonumber\\
&&~~
+ 
         66 (2 \barb (514 + 469 \barb) + 77 \barb_2) f 
+ 66 (278 + 819 \barb) f^2 + 
         15407 f^3\bigr) k_2^2\bigr\} \cos 2\theta 
+ \bigl\{7 \bigl(99 \barb^2 (7 + 3 f) + 66 \barb f (11 + 7 f) 
\nonumber\\
&&~~
+ 
         f^2 (297 + 205 f)\bigr) k_1^4 + \bigl(231 \barb (\barb (193 + 42 \barb) + 42 \barb_2) + 
         33 (\barb (1300 + 1057 \barb) + 154 \barb_2) f + 66 (206 + 553 \barb) f^2 
\nonumber\\
&&~~
+ 
         11914 f^3\bigr) k_1^2 k_2^2 + 
      7 (264 \barb f (5 + 3 f) + 198 \barb^2 (7 + 3 f) + 
         f^2 (363 + 265 f)) k_2^4\bigr\} \cos3\theta + 
   k_1 k_2 \bigl\{11 \bigl\{(189 \barb^2 (3 + f) 
\nonumber\\
&&~~
+ 6 \barb f (103 + 70 f) + 
         f^2 (255 + 217 f)\bigr) k_1^2 + (693 \barb^2 (13 + 6 f) + 
         66 \barb f (139 + 105 f) + f^2 (3597 + 2912 f)) k_2^2\bigr\} \cos4\theta 
\nonumber\\
&&~~+ 
   k_2^2 \bigl\{(1386 \barb^2 + 66 \barb f (29 + 21 f) + 7 f^2 (165 + 151 f)) k_1^2 + 
      21 f^2 (11 + 5 f) k_2^2\bigr\} \cos5\theta + 
   3 f^2 (44 + 35 f) k_1 k_2^3 \cos6\theta)
\bigr]
\nonumber
\\
\\
&&\hspace{-0.5cm}
C^2_{23}={f\over 97020 k_2 (k_1^2 + k_2^2 + 
     2 k_1 k_2 \cos\thetaot)^2}\bigl[
 -6 \bigl(77 \barb (\barb (107 + 28 \barb) - 35 \barb_2) 
+ 
      11 (\barb (652 + 623 \barb) - 91 \barb_2) f 
\nonumber\\
&&~~
+ 11 (173 + 574 \barb) f^2 + 
      1764 f^3\bigr) k_1^4 k_2 - 
   2 \bigl(8085 \barb (\barb - 4 \barb_2) + 66 (\barb (97 + 91 \barb) - 203 \barb_2) f + 
      99 (17 + 56 \barb) f^2 
\nonumber\\
&&~~
+ 1610 f^3\bigr) k_1^2 k_2^3 + 
   924 \barb_2 (7 \barb + 3 f) k_2^5 - 
   3 k_1 \bigl\{7 \bigl(616 \barb^3 + 33 \barb^2 (49 + 45 f) + 22 \barb f (61 + 53 f) + 
         f^2 (341 + 305 f)\bigr) k_1^4 
\nonumber\\
&&~~
+ 
      2 \bigl(77 \barb (\barb (173 + 56 \barb) - 182 \barb_2) + 
         22 (\barb (523 + 532 \barb) - 266 \barb_2) f + 22 (145 + 462 \barb) f^2 + 
         2940 f^3\bigr) k_1^2 k_2^2 
\nonumber\\
&&~~
- 
      3542 \barb_2 (7 \barb + 3 f) k_2^4\bigr\} \cos\theta - 
   3 k_2 \bigl\{2 \bigl(77 \barb (\barb (127 + 56 \barb) - 21 \barb_2) + 
         11 (746 \barb + 826 \barb^2 - 77 \barb_2) f 
\nonumber\\
&&~~
+ 121 (19 + 56 \barb) f^2 + 
         1918 f^3\bigr) k_1^4 + \bigl(154 \barb (34 \barb + 28 \barb^2 - 175 \barb_2) + 
         22 (\barb (232 + 259 \barb) - 539 \barb_2) f 
\nonumber\\
&&~~
+ 11 (137 + 420 \barb) f^2 + 
         1330 f^3\bigr) k_1^2 k_2^2 - 924 \barb_2 (7 \barb + 3 f) k_2^4\bigr\} \cos 2\theta - 
   k_1 \bigl\{7 \bigl(99 \barb^2 (7 + 3 f) + 66 \barb f (11 + 7 f) 
\nonumber\\
&&~~
+ 
         f^2 (297 + 205 f)\bigr)k_1^4 + 
      3 \bigl(77 \barb (\barb (143 + 56 \barb) - 84 \barb_2) + 
         77 (\barb (122 + 119 \barb) - 40 \barb_2) f + 22 (116 + 315 \barb) f^2 
\nonumber\\
&&~~
+ 
         1855 f^3\bigr) k_1^2 k_2^2 - 4158 \barb_2 (7 \barb + 3 f) k_2^4)\bigr\} \cos3\theta - 
   6 k_1^2 k_2 \bigl\{(693 \barb^2 (2 + f) + 22 \barb f (57 + 35 f) + 
         2 f^2 (209 + 119 f)) k_1^2 
\nonumber\\
&&~~
+ \bigl(231 \barb^2 (5 + 3 f) + 
         33 \barb (-49 \barb_2 + 2 f (13 + 7 f)) + 
         f (-693 \barb_2 + f (209 + 105 f))\bigr) k_2^2\bigr\} \cos4\theta - 
   3 (231 \barb^2 (5 + 3 f) 
\nonumber\\
&&~~
+ 66 \barb f (13 + 7 f) + 
      7 f^2 (44 + 15 f)) k_1^3 k_2^2 \cos5\theta - 
   99 f^2 k_1^2 k_2^3 \cos6\theta)
\bigr]
\\
&&\hspace{-0.5cm}
C^2_{31}={f\over 97020 k_1 (k_1^2 + k_2^2 + 
     2 k_1 k_2 \cos\thetaot)^2}
 \bigl[3696 \barb_2 (7 \barb + 3 f) k_1^5 - \bigl(1617 \barb (\barb (7 + 6 \barb) - 58 \barb_2) + 
      330 (\barb (37 + 49 \barb) 
\nonumber\\
&&~~
- 119 \barb_2) f + 33 (115 + 434 \barb) f^2 + 
      4270 f^3\bigr) k_1^3 k_2^2 - 
   3 \bigl(77 \barb (\barb (199 + 98 \barb) - 70 \barb_2) + 
      22 (\barb (607 + 749 \barb) - 91 \barb_2) f 
\nonumber\\
&&~~
+ 33 (111 + 392 \barb) f^2 + 
      3528 f^3\bigr)k_1 k_2^4 - 
   3 k_2 \bigl\{-4928 \barb_2 (7 \barb + 
         3 f) k_1^4 + \bigl(77 \barb (\barb (355 + 154 \barb) - 406 \barb_2) 
\nonumber\\
&&~~
+ 
         11 (\barb (2182 + 2443 \barb) - 1190 \barb_2) f + 11 (601 + 2044 \barb) f^2 + 
         6405 f^3\bigr) k_1^2 k_2^2 
+ 
      14 (308 \barb^3 + 693 \barb^2 (1 + f) 
\nonumber\\
&&~~
+ 14 f^2 (11 + 10 f) + 
         44 \barb f (13 + 12 f)) k_2^4\bigl\} \cos\theta 
   -3 k_1k_2^2 \bigl\{2 \bigl(77 \barb (\barb (55 + 7 \barb) - 133 \barb_2) + 
            11 (\barb (286 + 259 \barb) 
\nonumber\\
&&~~
- 413 \barb_2) f + 11 (69 + 203 \barb) f^2 + 
            595 f^3\bigr) k_1^2 + \bigl(154 \barb (5 \barb (26 + 7 \barb) - 21 \barb_2) + 
            22 (764 \barb + 700 \barb^2 - 77 \barb_2) f
\nonumber\\
&&~~
 + 33 (139 + 406 \barb) f^2 + 
            3836 f^3\bigr) k_2^2\bigr\} \cos 2\theta - 
      k_2^3 \bigl\{3 (1078 \barb (\barb (10 + \barb) - 3 \barb_2) + 
            11 (\barb (782 + 581 \barb) - 154 \barb_2) f 
\nonumber\\
&&~~
+ 22 (109 + 238 \barb) f^2 + 
            1435 f^3) k_1^2 + 
         7 (264 \barb f (5 + 3 f) + 198 \barb^2 (7 + 3 f) + 
            f^2 (363 + 265 f)) k_2^2\bigr\}\cos3\theta 
\nonumber\\
&&~~
- 
      3 k_1k_2^2 \bigl\{33 (21 \barb^2 + 22 \barb f + 9 f^2) k_1^2 + (693 \barb^2 (5 + 2 f) + 
            66 \barb f (47 + 21 f) + f^2 (957 + 476 f)) k_2^2\bigl\} \cos4\theta 
\nonumber\\
&&~~
- 
      3 k_2^3 \bigl\{33 (21 \barb^2 + 20 \barb f + 7 f^2) k_1^2 + 
         7 f^2 (11 + 5 f) k_2^2\bigl\} \cos5\theta - 
      99 f^2 k_1 k_2^4 \cos6\theta\bigr]
\end{eqnarray}

\begin{eqnarray}
&&C^4_{12}={f^2\over 5045040 k_1 k_2 (k_1^2 + k_2^2 + 
     2 k_1 k_2 \cos\thetaot)}
\bigl[ \bigl(2 (143 (\barb (2032 + 2023 \barb) + 112 \barb_2) + 13 (9616 + 28413 \barb) f 
\nonumber\\
&&~~
+ 
      128898 f^2\bigr) k_1^3 k_2 + 
   2 \bigl(249249 \barb^2 + 16016 \barb_2 + 26 \barb (8371 + 12019 f) + 
      9 f (10842 + 12397 f)\bigr) k_1 k_2^3 
\nonumber\\
&&~~
+ 
   2 \bigl\{56 \bigl(2717 \barb^2 + 78 \barb (33 + 46 f) + 
         f (1118 + 1215 f)\bigr) k_1^4 + \bigl(1001 (723 \barb + 750 \barb^2 + 80 \barb_2) + 
         13 (24517 
\nonumber\\
&&~~
+ 75719 \barb) f + 340767 f^2\bigr) k_1^2 k_2^2 + 
      7 (143 \barb (99 + 112 \barb) + 13 (483 + 1583 \barb) f + 
         7245 f^2) k_2^4\bigr\} \cos\theta 
\nonumber\\
&&~~
+ 
   k_1 k_2 \bigl\{104 \bigl(7777 \barb^2 + 11 \barb (688 + 987 f) + 
         6 (154 \barb_2 + 
            f (576 + 623 f))\bigr) k_1^2 + \bigl(572 (\barb (1401 + 1414 \barb) + 
            168 \barb_2) 
\nonumber\\
&&~~
+ 13 (27108 + 85631 \barb) f + 376677 f^2\bigr) k_2^2\bigr\} \cos 2\theta 
+ \bigl\{112 (143 \barb (6 + 5 \barb) + 26 (17 + 42 \barb) f + 
         465 f^2) k_1^4 
\nonumber\\
&&~~
+ \bigl(143 (\barb (5345 + 5138 \barb) + 672 \barb_2) + 
         13 (27553 + 81067 \barb) f + 381339 f^2\bigr) k_1^2 k_2^2 + 
      7 \bigr(143 \barb (151 + 160 \barb) 
\nonumber\\
&&~~
+ 13 (759 + 2419 \barb) f + 
         10665 f^2\bigr) k_2^4\bigr\} \cos3\theta + 
   2 k_1 k_2 ((143 \barb (592 + 525 \barb) + 13 (3440 + 8799 \barb) f + 
         53046 f^2) k_1^2 
\nonumber\\
&&~~
+ (143 \barb (982 + 805 \barb) + 
         26 (2715 + 6587 \barb) f + 70371 f^2) k_2^2) \cos4\theta + 
   k_2^2 \bigl\{\bigl(143 \barb (661 + 490 \barb) 
\nonumber\\
&&~~
+ 91 (579 + 1345 \barb) f + 
         66087 f^2\bigr) k_1^2 + 
      35 \bigl(91 \barb (11 + 5 f) + 3 f (169 + 115 f)\bigr) k_2^2\bigr\} \cos5\theta  
\nonumber\\
&&~~+   5 \bigl(91 \barb (44 + 35 f) + 3 f (676 + 805 f)\bigr) k_1 k_2^3 \cos6\theta\bigr]
\\
&&C^4_{23}={f^2\over 5045040 k_2 (k_1^2 + k_2^2 + 
     2 k_1 k_2 \cos\thetaot)^2}\bigl[
-2 \bigl(2288 (9 \barb (10 + 7 \barb) - 7 \barb_2) + 13 (6848 + 18781 \barb) f 
\nonumber\\
&&~~
+  91287 f^2\bigr) k_1^4 k_2 - 
   2 \bigl(32032 \barb^2 - 73073 \barb_2 + 30 f (715 + 602 f) + 
      52 \barb (935 + 938 f)\bigr) k_1^2 k_2^3 + 18018 \barb_2 k_2^5 
\nonumber\\
&&~~
- 
   2 k_1 \bigl\{6 \bigl(2717 \barb^2 + 78 \barb (33 + 46 f) + 
         f (1118 + 1215 f)\bigr) k_1^4 + \bigl(143 (\barb (2325 + 1792 \barb) - 
            896 \barb_2) 
\nonumber\\
&&~~
+ 13 (11253 + 31367 \barb) f + 
         152145 f^2\bigr) k_1^2 k_2^2 - 86086 \barb_2 k_2^4)\bigr\} \cos\theta - 
   k_2 \bigl\{8 (572 (2 \barb (61 + 77 \barb) - 21 \barb_2) 
\nonumber\\
&&~~
+ 13 (2536 + 8225 \barb) f + 
        37611 f^2) k_1^4 +\big(715 \barb (261 + 224 \barb) - 392392 \barb_2 
+ 
         13 (6303 + 20111 \barb) f 
\nonumber\\
&&~~
+ 97965 f^2\big) k_1^2 k_2^2 - 
      40040 \barb_2 k_2^4\bigl\} \cos 2\theta - 
   k_1 \bigl\{112 \big(143 \barb (6 + 5 \barb) + 26 (17 + 42 \barb) f + 
         465 f^2\big) k_1^4 
\nonumber\\
&&~~
+ \big(560560 \barb^2 - 256256 \barb_2 + 
         91 \barb (4125 + 6973 f) + 3 f (62543 + 74655 f)\big) k_1^2 k_2^2 - 
      270270 \barb_2 k_2^4\bigr\} \cos3\theta 
\nonumber\\
&&~~
- 
   2 k_2 \bigl\{(11440 \barb (8 + 7 \barb) + 39 (1056 + 2933 \barb) f + 
         40509 f^2) k_1^4 + (143 (8 \barb (33 + 70 \barb) - 805 \barb_2) 
\nonumber\\
&&~~
+ 
         26 (813 + 2870 \barb) f + 25200 f^2) k_1^2 k_2^2 - 
      35035 \barb_2 k_2^4\bigl\} \cos4\theta - 
   k_1 k_2^2 \bigl\{\big(80080 \barb^2 + 483 f (91 + 75 f) 
\nonumber\\
&&~~
+ 
         13 \barb (8679 + 9415 f)\big) k_1^2 - 70070 \barb_2 k_2^2\bigr\} \cos5\theta - 
   5 \big(455 \barb (11 + 7 f) + 3 f (507 + 245 f)\big) k_1^2 k_2^3 \cos6\theta)
\bigr]
\\
&&C^4_{31}={f^2\over 5045040 k_1 (k_1^2 +
      k_2^2 + 2 k_1 k_2 \cos\thetaot)^2}
\bigl[-32 \big(7007 \barb^2 + 5 f (403 + 567 f) + 
      13 \barb (275 + 602 f)\big) k_1^3 k_2^2 
\nonumber\\
&&~~
- 
   2 \big(143 \barb (1250 + 1393 \barb) + 78 (1003 + 3381 \barb) f + 
      91287 f^2\big) k_1 k_2^4 + 
   32032 \barb_2 k_1 (4 k_1^4 + 13 k_1^2 k_2^2 + k_2^4) 
\nonumber\\
&&~~
- 
   2 k_2 \bigl\{-256256 \barb_2 k_1^4 + 
      8 \big(143 (\barb (310 + 399 \barb) - 182 \barb_2) + 26 (793 + 2772 \barb) f + 
         25515 f^2\big) k_1^2 k_2^2 
\nonumber\\
&&~~
+ 
      7 \big(143 \barb (99 + 112 \barb) + 13 (483 + 1583 \barb) f + 
         7245 f^2\big) k_2^4\bigr\} \cos\theta 
   -k_1k_2^2  \bigl\{32 (143 (5 \barb (10 + 7 \barb) - 77 \barb_2) 
\nonumber\\
&&~~
+ 26 (107 + 329 \barb) f + 
            3045 f^2) k_1^2 + (143 (\barb (3839 + 4648 \barb) - 672 \barb_2) + 
            39 (6581 + 21840 \barb) f 
\nonumber\\
&&~~
+ 300888 f^2) k_2^2\bigr\} \cos 2\theta - 
      k_2^3 \bigl\{16 \big(715 \barb (34 + 21 \barb) - 6006 \barb_2 + 104 (101 + 252 \barb) f + 
            9765 f^2\big) k_1^2 
\nonumber\\
&&~~
+ 
         7 \big(143 \barb (151 + 160 \barb) + 13 (759 + 2419 \barb) f + 
            10665 f^2\big) k_2^2\bigr\} \cos3\theta - 
      6 k_1k_2^2  (208 (33 \barb + 17 f) k_1^2 
\nonumber\\
&&~~
+ 
         3 (715 \barb (18 + 7 \barb) + 26 (221 + 413 \barb) f + 
            4501 f^2) k_2^2) \cos4\theta - 
      k_2^3 \bigl\{1248 (44 \barb + 21 f) k_1^2 
\nonumber\\
&&~~
+ 
         35 (91 \barb (11 + 5 f) + 3 f (169 + 115 f)) k_2^2\bigr\} \cos5\theta 
- 195 (77 \barb + 39 f) k_1 k_2^4 \cos6\theta))
\bigr]
\end{eqnarray}

\end{appendix}
\end{document}